\documentclass[11pt]{article}
\usepackage{amsfonts}
\usepackage{amssymb}
\usepackage{latexsym}
\usepackage{graphicx}
\usepackage[english]{babel}
\input epsf

\renewcommand{\theequation}{\arabic{section}.\arabic{equation}}
\def\be{\begin{equation}}
\def\bea{\begin{eqnarray}}
\def\ee{\end{equation}}
\def\eea{\end{eqnarray}}
\def\c{\cosh\alpha}
\def\s{\sinh\alpha}
\def\r{\rightarrow}
\def\p{\partial}
\def\t{\tilde}

\begin{document}

\begin{flushright}
\end{flushright}
\vspace{20mm}
\begin{center}
{\LARGE  A Microscopic Model for the Black hole -- Black string  \\Phase Transition}\\
\vspace{18mm}
{\bf Borun D. Chowdhury}$^{a,}$\footnote{E-mail: {\tt borundev@mps.ohio-state.edu}.}, 
{\bf Stefano Giusto}$^{b,}$\footnote{E-mail: {\tt giusto@physics.utoronto.ca}.} 
{\bf and Samir D. Mathur}$^{a,}$\footnote{E-mail: {\tt mathur@mps.ohio-state.edu}.}\\
\vspace{8mm}
$^{a}$Department of Physics,\\ The Ohio State University,\\ Columbus,
Ohio, USA 43210\\
\vspace{4mm}

\vspace{4mm}
$^{b}$Department of Physics,\\ University of Toronto,\\ Toronto,
Ontario,  Canada M5S 1A7\\
\end{center}
\vspace{10mm}
\thispagestyle{empty}
\begin{abstract}

Computations in general relativity have  revealed an interesting phase diagram for the  black hole -- black string phase transition, with three different black objects present for a range of mass values. We can add charges to this system by `boosting' plus dualities; this makes only kinematic changes in the gravity computation but has the virtue of bringing the system into the near-extremal domain where a microscopic model can be conjectured. When the compactification radius is very large or very small then we get the microscopic models of 4+1 dimensional near-extremal holes and 3+1 dimensional near-extremal holes respectively (the latter is a uniform black string in 4+1 dimensions). We propose a simple model that interpolates between these limits and reproduces most of the features of the phase diagram. These results should help us understand how `fractionation' of branes  works in general situations.

\end{abstract}
\newpage
\setcounter{page}{1}
\renewcommand{\theequation}{\arabic{section}.\arabic{equation}}
\section{Introduction}\label{intr}
\setcounter{equation}{0}
Semiclassical physics gives us the entropy and Hawking radiation rates of black holes. In string theory we can understand extremal and near-extremal holes in terms of branes, and thereby reproduce the entropy and radiation from a microscopic description. 

In this paper we will consider another property of black holes that can be described in classical general relativity: the transition between black holes and black strings when a transverse circle is made large or small. We  suggest a simple microscopic picture of this transition, which will reproduce many of the broad features of the `phase diagram' of the system. 

\subsection{The `phase diagram'}

\begin{figure}[htbp] 
\vspace{.7truecm}
   \begin{center}
   \includegraphics[width=5in]{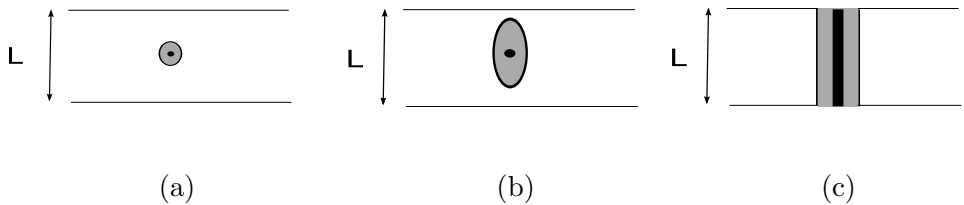} 
   \end{center}
   \vspace{.2truecm}
   \hspace{3.6truecm}(a)\hspace{4truecm}(b)\hspace{3.8truecm}(c)
   \caption{(a) A small black hole in a space with a compact circle of length $L$ \quad (b) The horizon distorts when the mass is increased so that its radius becomes comparable to $L$ \quad (c) At still larger masses we get a black string which wraps uniformly around the compact circle.}
   \label{fig:BH-BS_PT}
\end{figure}

In Fig.\ref{fig:BH-BS_PT}(a) we depict a small black hole in a spacetime with a compact transverse direction which we will call $z$; the length of this transverse circle is $L$. In Fig.\ref{fig:BH-BS_PT}(b) we depict a hole with a larger mass; the horizon now feels a significant distortion from the compactification. In Fig.\ref{fig:BH-BS_PT}(c) we have increased the mass still further; the black hole horizon has now turned into a `black string' horizon.

The black string has a nonzero `tension' ${\cal T}$ along the compact circle. The hole in Fig.\ref{fig:BH-BS_PT}(b) will also have some tension in this direction. When we take the hole to be very small, as in Fig.\ref{fig:BH-BS_PT}(a), the hole does not notice the compactification, and this tension becomes small too, vanishing for infinitesimal holes. The relevant parameter here is $r_0/L$, where $r_0$ is the radius of the horizon. When this parameter is small the effects of the compactification go away, and the tension becomes ignorable in the determination of the geometry of the hole.

\begin{figure}[t] 
   \begin{center}
   \includegraphics[width=3in]{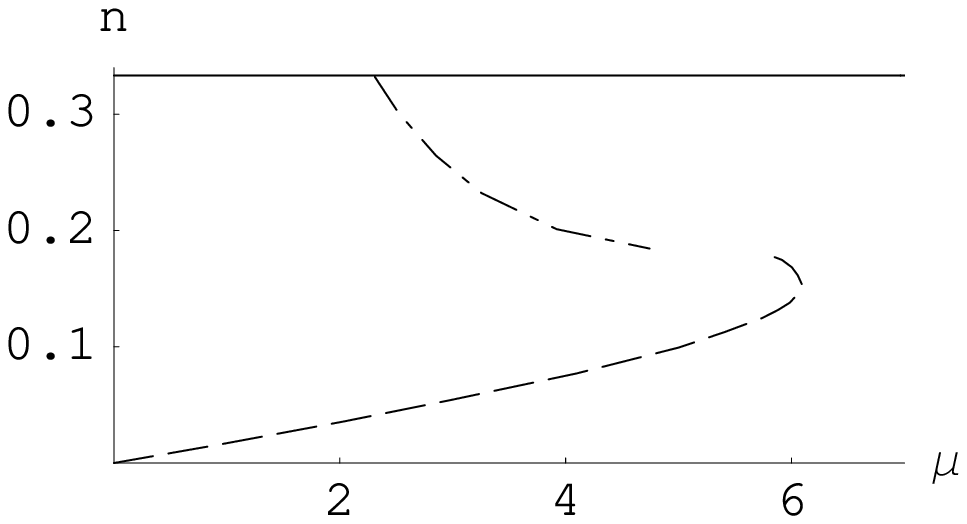} 
    \includegraphics[width=3in]{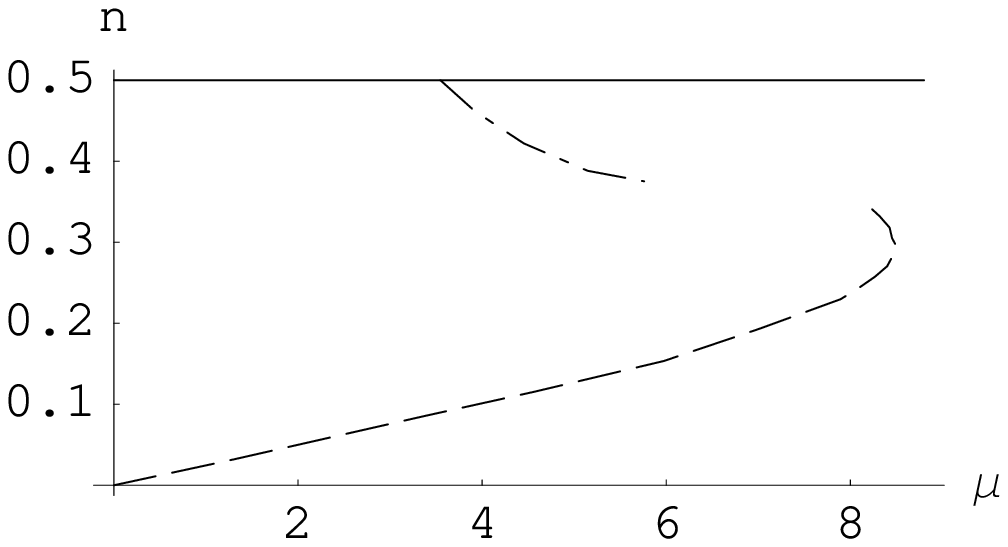} \\
    \vspace{.2truecm}
    \end{center}
    \hspace{4.5truecm} (a)\hspace{7truecm}(b)
     \caption{The phase diagram for  (a) 4 noncompact space directions and (b) 3 noncompact space directions. The vertical axis measures the dimensionless relative tension and  and the horizontal axis gives a dimensionless mass parameter. The solid line denotes the uniform black string, the dashed line denotes the black hole, and the dot-dashed line denotes the non-uniform black string.}
   \label{fig:nVSmu_gravity}
\end{figure}

Through a large number of studies, some analytical and some numerical, an interesting `phase diagram' has emerged for this system \cite{GL}-\cite{sahakian}. In Fig.\ref{fig:nVSmu_gravity}(a) we reproduce this diagram for the case of 5-D; i.e. 4 noncompact space  dimensions and the compact circle of length $L$. In Fig.\ref{fig:nVSmu_gravity}(b) we give the diagram for 4-D; i.e., 3 noncompact space directions and the compact circle. (Our interest will be in the 4-D case, but we give the diagram for the 5-D case as well since it is more complete -- numerical computations are easier in this case since the gravitational fields fall off faster at infinity.)

On the horizontal axis we have the mass, scaled by some constant and termed $\mu$. On the vertical axis we have a dimensionless tension $n={{\cal T} L\over M}$. For small $\mu$ we have just the `black hole' phase, with $n\r 0$ for $\mu\r 0$. At very large $\mu$ we have the `black string' phase, with $n$ independent of $\mu$. But the most striking   feature of these phase diagrams is that we do not see just a black hole and a black string; for a range of $\mu$ there is a third branch, called the `non-uniform black string'. 

\begin{figure}[ht] 
   \begin{center}
   \includegraphics[width=3in]{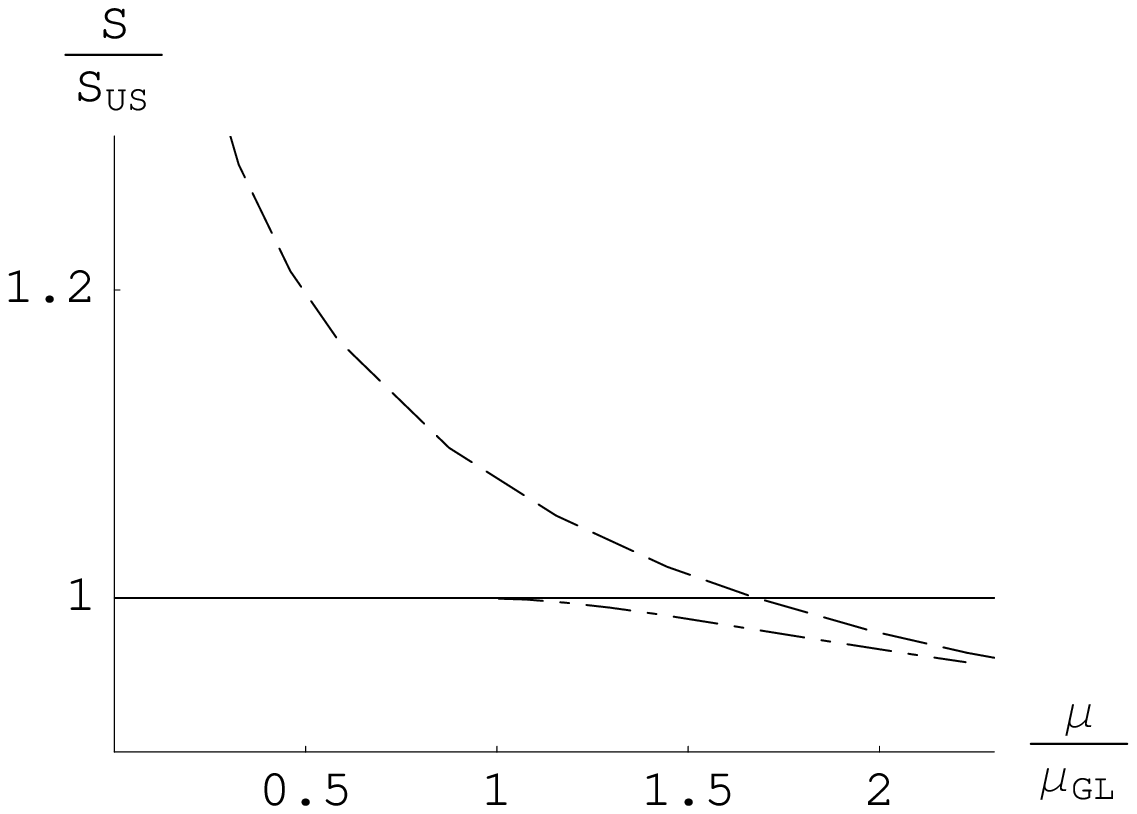} 
   \includegraphics[width=3in]{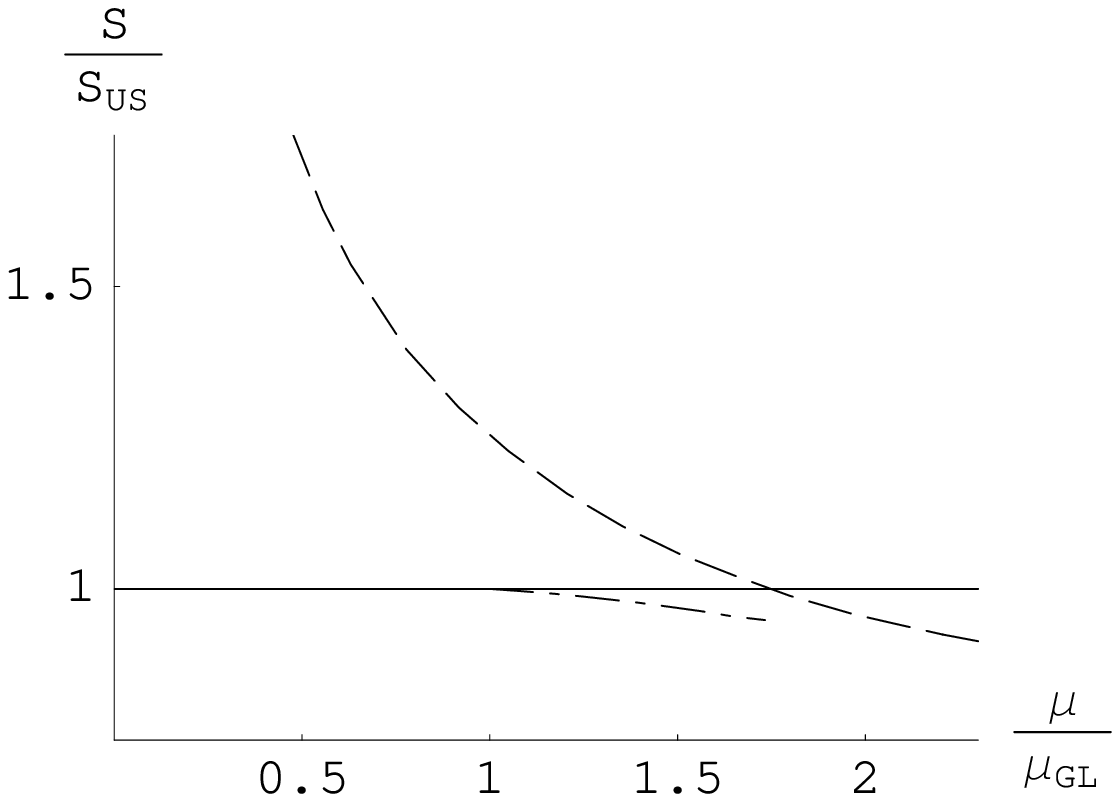} \\
   \vspace{.2truecm}
   \end{center}
   \hspace{4.5truecm}(a)\hspace{7truecm}(b)
    \caption{The entropies for (a) 4 noncompact space directions and (b) 3 noncompact space directions. The vertical axis gives the ratio of the entropy to the entropy of the uniform black string, and the horizontal axis gives a dimensional mass.
   The three kinds of lines label the three phases  in the same way as in  Fig.(\ref{fig:nVSmu_gravity})}
    \label{fig:sVSmu_gravity}
\end{figure}

In each of these phases we can compute the Bekenstein entropy from the area of the horizon. In Fig.\ref{fig:sVSmu_gravity}(a),(b) we reproduce the entropy graphs. The horizontal axis again gives the mass, and on the vertical axis we plot ${s\over s_\mathrm{us}}$, where $s_\mathrm{us}$ is the entropy of the uniform string.
Thus the uniform string branch just gives a horizontal line at height unity. The black hole branch is seen to have higher entropy at low mass, and lower entropy at high mass. The non-uniform string is seen to have an entropy that is always lower than the other two branches.

\subsection{The microscopic picture}

Our goal is to see if this phase diagram can be understood in terms of a microscopic description of black holes. The microscopic picture works best for extremal and near-extremal objects. This is because  supersymmetry prevents strong dependences on coupling and thus allows computations in terms of weakly interacting excitations. So our first goal is to map the phase diagram of neutral objects to the phase diagram for near-extremal objects. A detailed study of this map was made in \cite{harmarkoberscharged,hkor}, and we will use many of their results.

We will work in type IIB string theory. Consider the compactification
\be
M_{9,1}\r M_{4,1}\times T^4\times S^1
\ee
The compact circle of length $L$ will be chosen from the spatial directions in $M_{4,1}$, we will call it $\t S^1$.

We will add large D1 and D5 charges to the neutral object.
The addition of charges is done by `boosting' the system in the compact direction $S^1$; this gives a momentum charge P in that direction. Dualities convert this to a D1 charge along $S^1$. We boost again along $S^1$, and dualize to get a D5 charge on $T^4\times S^1$. 

First let $L\r\infty$, so we have a black object in 4+1 noncompact spacetime dimensions. This is the near extremal D1-D5 black hole studied in \cite{cm,ms}. In the microscopic description, the D1-D5 bound state produces an `effective string' winding along $S^1$ with total winding number $n_1n_5$. The  non-extremal energy of the system is carried by a dilute gas of $P, \bar P$ excitations along the $S^1$ direction, running along this effective string. The entropy of these excitations is given by
\be
S=2\pi\sqrt{n_1n_5}(\sqrt{n_p}+\sqrt{\bar n_p})
\label{en1}
\ee
With no net $P$ charge, we have $n_p=\bar n_p$, and the value of $n_p$ is given by the non-extremal energy.
So we have a microscopic picture of `small black holes', i.e. black holes where $r_0/L$ is small. 

Now let us imagine that $L$ is small. Then the effective compactification is
\be
M_{9,1}\r M_{3,1}\times T^4\times S^1\times \t S^1
\label{com2}
\ee
With this compactification we can make an extremal black hole with 4 kinds of charges: D1, D5 as above, P along $S^1$, and KK monopoles which have $\t S^1$ as their nontrivially fibered circle.  What we have however is two nonzero charges (D1, D5) plus nonextremal energy.  Note that the  4 kinds of charges can be permuted in all possible ways by dualities. Using this fact, we can write the entropy of this near-extremal system from the expression found in \cite{malda}, getting
\be
S=2\pi\sqrt{n_1n_5}(\sqrt{n_p}+\sqrt{\bar n_p})(\sqrt{n_k}+\sqrt{\bar n_k})
\label{en2}
\ee
If the net P, KK charges are zero then $n_p=\bar n_p$, $n_k=\bar n_k$, and the values of $n_p, n_k$ are determined by extremising $S$ over all possibilities for given total nonextremal energy. This entropy describes a black object that is uniformly smeared in all the compact directions of (\ref{com2}), so it is a `uniform black string' in the direction $\t S^1$.

One can write an expression similar to (\ref{en1}),(\ref{en2}) for the neutral hole as well \cite{hms, hlm}. But we have chosen to add charges to our system and look at the near-extremal systems (\ref{en1}),(\ref{en2}) because for these cases we also have a microscopic derivation of low energy Hawking radiation \cite{dm1,ms,km}. Thus not only is the count of states correct, their microscopic description must also be correct because they lead to the expected dynamics.

Let us return to our problem of interest. For large $L$, we have the `black hole phase', described by (\ref{en1}), and for small $L$ we have the `uniform black string' phase, described by (\ref{en2}). What happens when we reduce $L$ from large values to small values? We seek a microscopic picture of the intermediate $L$ regime, which will help us understand the three phases that emerge in the gravity computation; in particular we wish to understand the `non-uniform black string' in the microscopic picture.

\subsection{Outline of our computations}

We proceed in the following steps:

\bigskip

(A) A basic property of gravity solutions with a horizon  is the Smarr relation. This is a scaling relation that relates the tension ${\cal T}$ of the solution to the mass, charge and entropy. Since our microscopic model must describe a black object with a horizon, the Smarr relation must be satisfied by the model. We thus recall the Smarr relation and sketch its derivation.

\bigskip

(B) The gravity solution has a nontrivial metric in $M_{4,1}$. The compact directions $T^4\times S^1$ can be `added on trivially' to the neutral solution in $M_{4,1}$; i.e. the metric is flat and constant in these directions. From Einstein's equations one then finds that  the tension in these `trivial' directions must  be {\it nonzero} in general. Thus from a microscopic perspective, these directions are not so `trivial': one must wrap branes around these directions to obtain the required tension. The tension in these `trivial' directions is also important in getting the correct changes of physical parameters under the `boosting' which adds charge. We recall  the equations that relate old and new quantities upon   boosting; these were derived in \cite{harmarkoberscharged}, but we also supply a physical derivation
valid at weak gravitational coupling since this makes some of the physics more transparent.

\bigskip

(C) We next make a microscopic model that describes the transition from the black hole to the black string. The entropy (\ref{en1}) arises from momentum modes $n_p, \bar n_p$ that are fractionated in units of $1/n_1n_5$  by the presence of the D1, D5 branes \cite{dmfrac, maldasuss}.  The entropy (\ref{en2}) comes from two kinds of excitations (momentum and KK monopoles) which are again fractionated by the D1, D5 brane charges \cite{malda}. We make a simple model where a part $E_1$ of the nonextremal energy $E$ goes to creating entropy of the type (\ref{en1}) and the rest $E-E_1$ goes  to creating entropy of type (\ref{en2}). In this process we assume that a fraction $(1-x)$ of the 
`fractionating charge'  $n_1n_5$ goes to fractionating the  excitations used in (\ref{en1}), and the fraction $x$ goes to fractionating the excitations for (\ref{en2}).
We then extremise  over $E_1, x$ to find the preferred state of the microscopic system. 

We find that below some energy $E$ there is just one phase, the `black hole' phase, where all entropy is of type (\ref{en1}). But for $E_1$ greater than a certain critical value, there are three extrema of the entropy. One is a `black hole' with entropy entirely of type (\ref{en1}), one is a `uniform black string' which has entropy entirely of type (\ref{en2}), and a third exremum where the entropy is partly of type (\ref{en1}) and partly of type (\ref{en2}). This last extremum is a saddle point of the entropy, and  we identify it with the non-uniform black string. The tension and entropy graphs for this very simple microscopic model are given in section \ref{microsection}, in Fig.\ref{fig:4DnVSmu_micro} and Fig.\ref{fig:4DsVSmu_micro} respectively. 

\bigskip

(D) While this simple model reproduces the three phases seen in the gravity computations, we notice one simple qualitative difference between the microscopic and gravity computations: the microscopic model has a vanishing tension in the black hole phase, while the gravity description gives a tension that rises from zero as the mass is increased from zero. We argue that this tension should be found in the microscopic picture if we include the `twist operator interaction' in the microscopic CFT, something we have not done in our leading order computation. 

An indirect approach to computing the tension of the black hole branch was taken in \cite{costaperry,hkor}, and we extend this approach here. One computes the tension from the gravity description, and uses this to read off the changes in excitations in the microscopic description. In the microscopic description we can imagine integrating out the effects of the KK monopole pairs (which are only weakly excited at large $L$) to obtain an `effective interaction' between the $P, \bar P$ excitations of the black hole. We first use the leading order gravity result to find this interaction to leading order in the microscopic picture. Then we use analyticity in the microscopic variables to predict the tension to next order on the gravity side, and observe agreement with the known gravity computation at that order.

We thus obtain some understanding of tension in the black hole phase. We note some agreements and also some disagreements between the gravity phase diagram and the microscopic diagram. We hope to return in future work to exploring further the  interactions in the microscopic picture in order to  obtain a more accurate phase diagram.

\section{Some basic relations}
\setcounter{equation}{0}

\subsection{The Smarr relation}

We recall the Smarr relation \cite{harmarkobersII}, and sketch a derivation; derivations along these lines can be found for example in \cite{kolsorpirI}. 

The full theory of quantum gravity has an inbuilt length scale -- the planck length $l_p$. But $l_p$ involves $\hbar$, and the classical theory does not have this length scale. The action $\int \sqrt{-g}~ R$ does not have any natural length scale, so if we have one solution to the vacuum equations, we can obtain another by uniformly scaling up all lengths by the same factor. Let the length of the compact direction $z$ change as
\be
L~\r~ L+dL
\label{zero}
\ee
Consider a black hole in a spacetime with one compact direction $z$ of length $L$. Let the number of noncompact spacetime dimensions be $d$; thus including the compact circle $z$ the total spacetime dimension is $d+1$. In this dimension the mass $M$ scales as $L^{d-2}$. (The mass appears in the metric through the combination $G_N^{(d+1)}M$, which has units $L^{d-2}$; $G_N^{(d+1)}$ is held
fixed, and thus $M$ scales as $L^{d-2}$.) Thus  we will get
\be
M~\r ~M\Bigl(1+{dL\over L}\Bigr)^{d-2}, ~~~dM=M(d-2){dL\over L}
\label{one}
\ee
The entropy is proportional to the surface area of the horizon. The horizon is a $d-1$ dimensional surface, so it will scale as
\be
S~\r ~S\Bigl(1+{dL\over L}\Bigr)^{d-1}, ~~~dS=S(d-1){dL\over L}
\label{two}
\ee
The charge $Q$ appears in the metric the same way that the mass does, so we have
\be
Q~\r ~Q\Bigl(1+{dL\over L}\Bigr)^{d-2}, ~~~dQ=Q(d-2){dL\over L}
\label{three}
\ee

Now assume that the solution satisfies the first law of thermodynamics (this will be true for all black objects)
\be
dE=TdS+{\cal T} dL+\phi\, dQ
\ee
where ${\cal T}$ is the `tension' in the $z$ direction and $\phi$ the potential at the horizon. This law will be satisfied in particular for the changes (\ref{zero})-(\ref{three}), since these changes lead from one valid solution to another. We thus get
\be
E(d-2)=TS(d-1)+{\cal T} L+\phi \,Q(d-2)
\ee
or
\be
TS={(d-2)(E-\phi\,Q)-{\cal T} L\over d -1}
\label{smarrnewp}
\ee
This is the Smarr relation.

\subsection{Behavior under boosting}

\label{boostsec}
Consider the metric of a black object, by which we mean any solution of gravity with a single connected horizon.  We assume that the solution is neutral, i.e. it carries no charge. The metric is of the form
\be
ds^2=-Udt^2+ds_B^2
\label{original}
\ee
where $U$ is independent of $t$ and the ``base metric'' $ds_B^2$  involves only the spatial coordinates. The horizon is at the location $U=0$.
We can add one or more `trivial directions' to this metric; we single out a special one called $y$ for later use:
\be
ds^2=-Udt^2+ds_B^2+dy^2+dz_adz_a
\label{neutral}
\ee
We will think of the added directions as compact circles. In particular let $0\le y<L_y$.

We can start with the solution (\ref{neutral}) and add charges to it by boosting plus dualities. Thus writing
\be
t=t'\c +y'\s, ~~~y=t'\s + y'\c
\label{boost}
\ee
we get
\be
ds^2=H^{-1}[-Udt'^2+H^2 (dy'-A_{t'}\, dt')^2]+ds_B^2+dz_adz_a
\label{charged}
\ee
where
\be
H=1+(1-U) \sinh^2\alpha, ~~~~A_{t'}=(H^{-1}-1) \coth\alpha
\ee

This process adds `momentum charge $P_y$'. We can change this to a different kind of charge by S,T dualities, and add further charges by boosting. We will write down a 2-charge solution in the Appendix, but for now we will look at the simple case having just $P_y$ charge. We can extract the following properties of the boosted solution; the derivations can be seen from the 2-charge case discussed in the Appendix.

Let the neutral solution (\ref{neutral}) have mass $M$, entropy $S$, and temperature $T$. (The temperature is extracted from the surface gravity at the horizon in the usual way.) The part $ds_B^2$ describes a $(d-1)$ dimensional noncompact space and a compact direction $z$, and we will let ${\cal T}$ denote the tension in the $z$ direction.

What are the properties of the charged solution (\ref{charged})? Note that $y$ was a compact direction, and we do not have the boost symmetry  (\ref{boost}) for compact $y$. But since the classical solution is homogeneous in $y$, we can lift the solution to the covering space of $y$, apply the boost, and then `recompactify' the new coordinate $y'$. Since the boosted solution is again homogeneous in $y'$, we can pick any coordinate length to compactify $y'$. Let us choose
$0\le y'<L_y$, thus keeping the length of the compact circle the same as in the neutral solution. Then we find the following \cite{harmarkoberscharged}:

\bigskip
(a) The mass $M'$ of the charged solution (\ref{charged}) is given by
\be
M'=M\Bigl(1+{d-2-n\over d-1} \sinh^2\alpha\Bigr)
\label{massgr}
\ee
where the relative tension $n$ is defined as
\be
n\equiv {L\mathcal{T}\over M}
\label{relative}
\ee

(b) The tension ${\cal T}'$ of the charged solution is the same as that of the neutral solution
\be
{\cal T}'={\cal T}
\label{tensiongr}
\ee

(c) The charge of the solution is
\be
Q=M {d-2-n\over d-1} \s\c
\label{chargegr}
\ee

(d) The entropy $S'$ of the charged solution is
\be
S'=S \c
\label{entropygr}
\ee

(e) The temperature $T'$ of the charged solution is
\be
T'={T\over \c}
\ee
Note that
\be
T'S'=TS
\ee

(f) The electric potential at the horizon is
\be
\phi=\tanh\alpha
\ee

\subsection{Understanding the `boost relations'}

The above relations (a)-(f) between the neutral and charged solutions are straightforward to derive in the gravity description of black objects. But our goal is to understand black objects from a microscopic point of view. From this `matter' point of view these relations appear somewhat nontrivial. They tell us the thermodynamic properties of a charged system once we know the properties of the neutral system. But abstractly speaking, there is no general relation between charged and neutral systems. So we must be looking at very special systems, and our microscopic model must be made to reproduce the properties of such special systems. To understand these properties better, we now rederive the relations of the above section from a  microscopic perspective.

\subsection{The matter stress tensor}

Consider the neutral system. Let the total spacetime dimension be $D$. The Einstein equations are 
\be
R_{AB}-{1\over 2} g_{AB} R=8\pi G\, T_{AB}
\ee
We can rewrite these as
\be
R_{AB}=8\pi G\,\t T_{AB}
\ee
where
\be
\t T_{AB}=T_{AB}-{1\over D-2} g_{AB} T
\ee
We will now assume that the gravitational field is weak. This will not be the case in general for the systems we finally consider, but  we are only trying to get a physical feeling for the relations (a)-(f) which are already known to be correct, and this weak-field assumption will allow us to extract some basic ideas in a simple way. Writing $g_{AB}=\eta_{AB}+h_{AB}$ we get (in the gauge $h_{AB,}{}^B=0$)
\be
-{1\over 2}\square h_{AB}-{1\over 2} h_{,AB}=8\pi G\,\t T_{AB}
\label{final}
\ee

Recall that we had started with a metric of the form (\ref{original}) and added `trivial' directions to obtain the metric (\ref{neutral}). It seems from the gravity point of view that these extra directions play no role in the physics of the black object. 
But from a microscopic viewpoint we see a different story. Let the weak gravity system (\ref{final})  describe this situation where we have trivially added directions $y, z_a$, and the metric is nontrivial in the noncompact directions $x_\mu$ and the compact direction $z$.  Consider one of the `trivial' directions, say $y$. First, we have {\it homogeneity} in $y$, so fields do not depend on $y$. Thus in (\ref{final}) we will get $h_{,yy}=0$. Second, we have {\it uniformity} of the $y$ circle, which means that the size of this circle does not change from place to  place. This gives $h_{yy}=0$.  Eq. (\ref{final}) then gives $\t T_{yy}=0$, which is
\be
T_{yy}-{1\over D-2} T=T_{yy}-{1\over D-2}[-T_{00}+T_{zz}+T_{yy}+T_{z_az_a}]=0
\label{basic}
\ee
Note that $T\ne 0$ in general, so $T_{yy}\ne 0$ in general. This means that our microscopic model {\it cannot} consist of branes that extend only in the directions $x_i, z$ while ignoring the trivial directions $y, z_a$. Branes have a tension only along their worldvolumes, and such a model would give $T_{yy}=0$. Instead, we must have at least one kind of brane wrapping each of the `trivial' directions. As an example consider the Schwarzschild solution in 4+1 noncompact dimensions. We can add the other 5 directions of string theory as `trivial directions' in the gravity solution. But a microscopic model for this system can be written in terms of branes and antibranes of type D1, D5, P \cite{hms}, and these objects are seen to wrap around  the 5 `trivial' directions.

The mass of the neutral system is
\be
M=\int [dx_i dz dy dz_a]\, T_{00}
\ee
The tension ${\cal T}$ along the $z$ direction is
\be
{\cal T}=-{\int [dx_i dz dy dz_a]\, T_{zz}\over L}
\ee
First note that under the boost (\ref{boost}) we have $T'_{zz}=T_{zz}$, so the tension remains unchanged
\be
{\cal T}'={\cal T}
\ee
which is the relation (\ref{tensiongr}).
Next, note that
\be
M'=\int [dx_i dz dy dz_a]\, T'_{00}=\int [dx_i dz dy dz_a]\, (T_{00}\cosh^2\alpha+T_{yy} \sinh^2\alpha)
\label{mass}
\ee
In (\ref{basic}) the `trivial' directions $y, z_a$ are all on the same footing, so $T_{yy}=T_{z_az_a}$. We then find
\be
T_{yy}=-{1\over d-1}(T_{00}-T_{zz})
\label{tyy}
\ee
Substituting this in (\ref{mass}) we find
\be
M'=M \cosh^2\alpha -{1\over d-1}(M+{\cal T} L)\sinh^2\alpha =M+M {d-2 -n\over d-1}
\sinh^2\alpha
\label{massnew}
\ee
where we have used the definition $n={\cal T} L/M$. Thus we reproduce (\ref{massgr}).

The charge $P_y$ is
\be
Q=\int [dx_i dz dy dz_a]\, T'_{0y}=\int [dx_i dz dy dz_a]\, (T_{00} \s\c+T_{yy} \s\c)
\ee
Using (\ref{tyy}) we find
\be
Q={(d-2)M -{\cal T} L\over d-1} \s\c=M {d-2 -n\over d-1} \s \c
\label{chargenew}
\ee
which agrees with (\ref{chargegr}).
\subsubsection{Entropy}

Now consider the entropy $S$. The neutral system was homogeneous in the direction $y$; we used this fact in lifting to the noncompact covering space before boosting. The classical entropy is  linear in the coordinate length of the $y$ circle, since the area of the horizon is proportional to this length. Thus the entropy of the microscopic model must be extensive in the length $L_y$.

Under boosting along $y$ we would expect that the length of the $y$ direction would contract by a factor $\gamma={1\over \sqrt{1-v^2}}=\c$; thus it would fit in a length $L'_y=L_y/\c$. We could have chosen any length of recompactification after boosting, but our choice was to keep $L_y$ unchanged. This means that we have considered a `larger amount of matter' in the charged case -- larger by a factor $\c$. This accounts for the relation
\be
S'=S \c
\label{entropynew}
\ee
\subsubsection{The relation $\phi=\tanh\alpha$}
The potential of the charged system is defined as
\be
\phi=\Bigl({\p M'\over \p Q}\Bigr)_{S'}
\ee
Setting $dS'=0$ in (\ref{entropynew}) we find
\be
S \sinh\alpha\, d\alpha+{dS\over dM}\cosh\alpha\,dM=0
\label{alpha}
\ee
Then from (\ref{massnew}) we find (using (\ref{alpha}) to eliminate $d\alpha$)
\be
dM'=dM\,\Bigl[1-2 {dS\over dM}{M\over S}{d-2-n\over d-1}\cosh^2\alpha +
{d-2-n -M {dn\over dM}\over d-1} \sinh^2\alpha\Bigr]
\ee
From (\ref{chargenew}) we find
\be
dQ=dM\,\Bigl[-\cosh 2\alpha{\c\over\s}{dS\over dM}{M\over S}{d-2-n\over d-1} 
+ {d-2 -n -M {dn\over dM}\over d-1} \s\c\Bigr]
\ee
As it stands, the ration ${dM'\over dQ}$ does not simplify. But now we use the Smarr relation for the neutral system
\be
TS=M {d-2-n\over d-1} \quad \Rightarrow\quad
{dS\over dM}{M\over S}{d-2-n\over d-1}=1
\label{smarrnew}
\ee
Then we find
\be
\phi={dM'\over dQ}=\tanh\alpha
\ee
\subsubsection{The relation $T'S'=TS$}
We have
\be
T'=\Bigl({d M'\over dS'}\Bigr)_Q
\ee
Setting $dQ=0$ in (\ref{chargenew}) we find
\be
d\alpha=-{dM\over M}\,{d-2-n-M{dn\over dM}\over d-2-n}\,
{\s\c\over \sinh^2\alpha+\cosh^2\alpha}
\ee
Using this we find
\be
dM'=dM\,\Bigl[1-{d-2-n-M{dn\over dM}\over d-1} {\sinh^2\alpha\over \sinh^2\alpha+ \cosh^2\alpha}
\Bigr]
\ee
\be
dS'={dM\over M}\,\Bigl[ M {d S\over dM}- S\, {d-2-n -M {dn\over dM}\over d-2-n}
{\sinh^2\alpha\over \sinh^2\alpha+\cosh^2\alpha}\Bigr]\c
\ee
As it stands the expression
\be
T'S'={dM'\over dS'} S'
\ee
does not simplify. But if we use the Smarr relation (\ref{smarrnew}) we find
\be
T'S'=TS
\ee

\subsection{The near extremal system}

Let us now note some results that will be useful in our study of the near extremal system that we will use. We will for the most part use D1,D5 charges, but will also have occasion to consider the system with three nonzero charges -- D1,D5,P.

\subsubsection{Two charges}
\label{2chargesection}
Let us take the solution (\ref{neutral}) and add D1-D5 charges. We find the solution (we give the derivation in the Appendix)
\bea
ds^2_E &\!\!=\!\!& H_1^{1/4}H_5^{-1/4}[H_1^{-1}(- U dt^2+ dy^2) + H_5 ds^2_B +  dz_a dz_a]\nonumber\\
 G^{(3)}&\!\!=\!\!&-U^{-1/2} \partial_r H_5 \coth\alpha_5 \,\star_B dr-\partial_r H_1^{-1}\coth\alpha_1\, dr\wedge dt\wedge dy\nonumber\\
e^{2\Phi}&\!\!=\!\!&{H_1\over H_5}\,,\quad H_1 = 1+(1-U)\sinh^2\alpha_1\,,\quad H_5 = 1+(1-U)\sinh^2\alpha_5
\label{chargedE}
\eea
(Here $ds_E^2$ is the Einstein metric.) In this case the analog of (\ref{massnew}) is
\be
M'=M\Bigl[1+{2-n\over 3}(\sinh^2\alpha_1+\sinh^2\alpha_5)\Bigr]
\label{mprime}
\ee
Since we will be working close to extremality, let us compute the energy above extremality
\be
E\equiv M'-M_{ex}=M'-Q_1-Q_5
\label{mex}
\ee
The charges $Q_1, Q_5$ are given by the analog of (\ref{chargenew})
\be
Q_1=M{2-n\over 3}\sinh\alpha_1\cosh\alpha_1, ~~~~~Q_5=M{2-n\over 3}\sinh \alpha_5\cosh\alpha_5
\label{q1q5charges}
\ee
In the near extremal limit we have $\alpha_1, \alpha_5\gg1$, so that
\be
\sinh\alpha_i\cosh\alpha_i\approx\sinh^2\alpha_i+{1\over 2}, ~~~~i=1,5
\ee
Using this in (\ref{mex}) we get
\be
E\approx M {1+n\over 3}
\label{m2e}
\ee
For the near extremal case it is convenient to define the relative tension in a manner
similar to the definition (\ref{relative}) but with the mass  replaced by the mass above extremality \cite{harmarkoberscharged}
\be
r\equiv {L{\cal T}'\over E}
\label{defr}
\ee
From (\ref{tensiongr}) and (\ref{m2e}) we find
\be
r={L{\cal T}\over E}={L{\cal T}\over M}{M\over E}={3n\over 1+n}
\label{n2r}
\ee

We can similarly find the transformation law for the entropy. With two charges the analogue of (\ref{entropygr}) 
is
\be
S'=S \cosh\alpha_1 \cosh\alpha_5
\label{entropy2ch}
\ee
For later use let us define a `rescaled entropy' as
\be
\hat{S}\equiv {S'\over\sqrt{Q_1 Q_5}}
\label{rescaledsgrav}
\ee
From (\ref{entropy2ch}) and (\ref{q1q5charges}) we find
\be
\hat{S}={S\over M} {3\over 2-n}{\cosh\alpha_1\cosh\alpha_5\over \sqrt{\sinh\alpha_1 \cosh\alpha_1 \sinh\alpha_5 \cosh\alpha_5}}
\ee
In the limit of large charges $\alpha_1,\alpha_5 \gg 1$ we get
\be
\hat S={S\over M} {3\over 2-n}
\label{s2shat}
\ee
 
\subsubsection{Three charges}

We will also have occasion to turn on 3 charges, D1, D5, P. We will have the D1, D5 charges large as above, but the P charge will be arbitrary. The relation (\ref{mprime}) now becomes
\be
M'=M\Bigl[1+{2-n\over 3}(\sinh^2\alpha_1+\sinh^2\alpha_5+\sinh^2\alpha_p)\Bigr]
\label{mprimep}
\ee
and we have
\be
Q_p=M{2-n\over 3}\sinh\alpha_p\cosh\alpha_p
\label{qpcharge}
\ee
We define the energy above extremality $E$ again as the energy above the mass of the large charges D1, D5, and find
\be
E\equiv M'-Q_1-Q_5=M {1+n\over 3}+M {2-n\over 3}  \sinh^2\alpha_p
\label{enew}
\ee
For later use, it will be helpful to write this relation in another way. For the neutral system (i.e. before any charges were added) we have the Smarr relation
\be
TS=M {2-n\over 3}
\ee
This implies
\be
M={3\over 2} TS+{nM\over 2}={3\over 2} TS+{{\cal T}L\over 2}
\ee
Substituting this in (\ref{enew}) we find
\be
E={TS\over 2}\cosh 2\alpha_p +{{\cal T}L\over 2}
\label{erelation}
\ee

\section{Phase Transition: Microscopic Model}
\setcounter{equation}{0}
\label{microsection}

We are looking at type IIB string theory with the compactification $M_{9,1}\r M_{4,1}\times T^4\times S^1\times \t S^1$.
On these 6 compact directions we can wrap 4 mutually BPS charges:

\medskip

(i)  D5 branes wrapped on $T^4\times S^1$.

(ii) D1 branes wrapped on $S^1$.

(iii) Momentum modes P that along along $S^1$.

(iv) KK-monopoles that have $\t S^1$ as their nontrivial fiber, and that extend uniformly along $T^4\times S^1$.

\medskip

The four charges above can be permuted in all possible ways among themselves by S,T dualities. For this reason we will
often list the number of these charges as $n_1, n_2, n_3, n_4$, instead of writing $n_5, n_1, n_p, n_k$.

Work on black hole microscopics has given an understanding of the entropy of black objects for various compactifications.
Let us recall these results.

First let the length $L$ of $\t S^1$ be infinite, so that we have only the compactification $T^4\times S^1$.
Then the mass of the KK charge becomes infinite, and it drops out of the computations and the entropy is an expression
in the three remaining charges $n_1, n_2, n_3$.
The entropy for all such cases can be derived from the following abstract expression \cite{hms}
\be
S=2\pi(\sqrt{n_1}+\sqrt{\bar n_1})(\sqrt{n_2}+\sqrt{\bar n_2})(\sqrt{n_3}+\sqrt{\bar n_3})
\label{threec}
\ee
There are 6 variables on the right hand side. We have to vary these variables to achieve a maximum of $S$, subject to the constraints that we are given the net values of the 3 types of charges
\be
\hat n_i=n_i-\bar n_i, ~~~i=1,2,3
\label{threed}
\ee
and the total energy
\be
M=\sum_{i=1}^3 m_i (n_i+\bar n_i)
\label{threee}
\ee 
where $m_i$ is the mass of each quantum of type $i$. Roughly speaking,  extremising
$S$ tells us that we create pairs of those charges that are lighter, and that have {\it smaller} net charge values;
i.e., we create pairs of the charges that we {\it don't} have. In particular, if some charge $\hat n_i\r \infty$, then the nonextremal energy will  not create pairs for this charge.

Let us list those cases that will be relevant to us in what follows:

\medskip

(a) Let there be 2 large charges $n_1, n_2\r \infty$, $\hat n_3=0$, and a small amount of nonextremal energy $E$. Then we find
\be
n_3=\bar n_3={E\over 2 m_3}
\ee
and
\be
S=2\pi\sqrt{n_1n_2}(\sqrt{n_3}+\sqrt{\bar n_3})=4\pi\sqrt{n_1n_2}\sqrt{E\over 2m_3}
\label{en1p}
\ee

(b) Let there be one large charge $n_1\r \infty$, $\hat n_2=\hat n_3=0$, and a small amount of nonextremal energy $E$. Then we have to extremise
\be
S=2\pi\sqrt{n_1}(\sqrt{n_2}+\sqrt{\bar n_2})(\sqrt{n_3}+\sqrt{\bar n_3})=8\pi\sqrt{n_1}\sqrt{n_2}\sqrt{n_3}
\label{en23}
\ee
subject to
\be
E=2m_2n_2+2 m_3n_3
\ee
We find
\be
n_2=\bar n_2={E\over 4 m_2}, ~~~~~n_3=\bar n_3={E\over 4 m_3}
\ee 
and
\be
S=2\pi\sqrt{n_1} {E\over \sqrt{m_2m_3}}
\label{en24}
\ee

\bigskip

Now let $\t S^1$ be finite, so that all 4 charges are relevant. The analog of (\ref{threec})-(\ref{threee}) are \cite{hlm}
\be
S=2\pi(\sqrt{n_1}+\sqrt{\bar n_1})(\sqrt{n_2}+\sqrt{\bar n_2})(\sqrt{n_3}+\sqrt{\bar n_3})(\sqrt{n_4}+\sqrt{\bar n_4})
\label{threecq}
\ee
\be
\hat n_i=n_i-\bar n_i, ~~~i=1,2,3,4
\label{threedq}
\ee
\be
M=\sum_{i=1}^4 m_i (n_i+\bar n_i)
\label{threeeq}
\ee 

The analogues of cases (a), (b) above are

\bigskip

(a') We have $\hat n_1, \hat n_2, \hat n_3\to \infty$, $\hat n_4=0$. We get
\be
n_4=\bar n_4={E\over 2 m_4}
\ee
and
\be
S=4\pi\sqrt{n_1n_2n_3}\,\sqrt{E\over 2m_4}
\label{formulap}
\ee

(b') We have $\hat n_1, \hat n_2\r\infty$, $\hat n_3=\hat n_4=0$. We get
\be
n_3=\bar n_3={E\over 4 m_3}, ~~~~~n_4=\bar n_4={E\over 4 m_4}
\label{formula}
\ee 
and
\be
S=2\pi \sqrt{n_1n_2}\, {E\over \sqrt{m_3m_4}}
\label{en2p}
\ee

\bigskip

Consider any of these expressions, say  (\ref{en1p}). We can regard the entropy as arising from 
`fractional' $n_p, \bar n_p$ pairs.  We have $n_1n_5n_p$ units of fractional $P$ excitations, and $n_1n_5\bar n_p$ units of fractional $\bar P$ excitations. Our interest is in understanding more about how fractionation works. We will therefore examine different possible fractionations below, and see how well they fit the physics of the black hole -- black string transition.

\subsection{Modeling the phase transition}
\label{microsectionmodel}
Let us now come to the main computations of this paper.  We have seen that in the limit of large $L$ and small $L$ the entropy of the system is given by (\ref{en1}) and (\ref{en2}) respectively. We now wish to make a simple model for the transition between these two possibilities. We have two large charges D1, D5 (with integer values $n_1, n_5$ respectively); we have added these to the system to make the system `near-extremal'. On the gravity side the addition of charges induced a simple map between neutral and charged systems, so no generality was lost by looking at this near-extremal case.

Let the energy above extremality be $E$. We let a part $E_1$ of this energy go to exciting $P, \bar P$ pairs. If these were the only excitations present then they would  describe a `black hole' in 4+1 noncompact dimensions, and give an entropy of type (\ref{en1}), which we have rewritten in the form (\ref{en1p}) above. The rest of the energy $E-E_1$ will go to entropy of the form (\ref{en2}); we have rewritten this entropy in the form (\ref{en2p}) above.  If all the energy was in this latter form we would get a uniform black string. The important question now is what happens to the charges $n_1, n_5$. There are three possibilities:

\bigskip

(a) A fraction $(1-x)$ of the charges $n_1$ go to the entropy of type (\ref{en1p}), and the rest $x$ go to the entropy (\ref{en2p}); similarly a fraction $(1-x)$ of the charges $n_5$ contribute to (\ref{en1p}), and the rest $x$ contribute to (\ref{en2p}). (We take the fractions $x$ equal for the two charges by the symmetry between these charges, but we can consider different fractions too.) Thus we have two disjoint systems, and the `fractionation' available for the two types of entropy are $x^2n_1n_5$ and $(1-x)^2 n_1n_5$ respectively. 

\medskip

(b) A fraction $(1-x)$ of the product $n_1n_2$ goes to the entropy (\ref{en1p}) and a fraction $x$ to the entropy (\ref{en2p}).
Thus the D1 and D5 branes make a bound state with $n_1n_5$ effective degrees of freedom, and it is these effective degrees of freedom that get partitioned into the two subsystems. It is useful to define
\be
N=n_1n_5
\ee
and the partitioning  gives $n_1'n_5'=N-\delta N$ to the entropy (\ref{en1p}) and $n_1''n_5''=\delta N$ to the entropy (\ref{en2p}).

\medskip

(c) All the charges $n_1$ and all the charges $n_5$ contribute to each kind of entropy. Thus the full product $n_1n_5$ appears in each of the expressions of type (\ref{en1p}) and (\ref{en2p}).

\bigskip

It turns out that possibility (b) gives a phase diagram that resembles the gravity computation, and we will adopt this choice below. We will discuss what happens with possibilities (a),(c) at the end of the section.

\begin{figure}[htbp] 
\vspace{.7truecm}
   \begin{center}
   \includegraphics[width=2in]{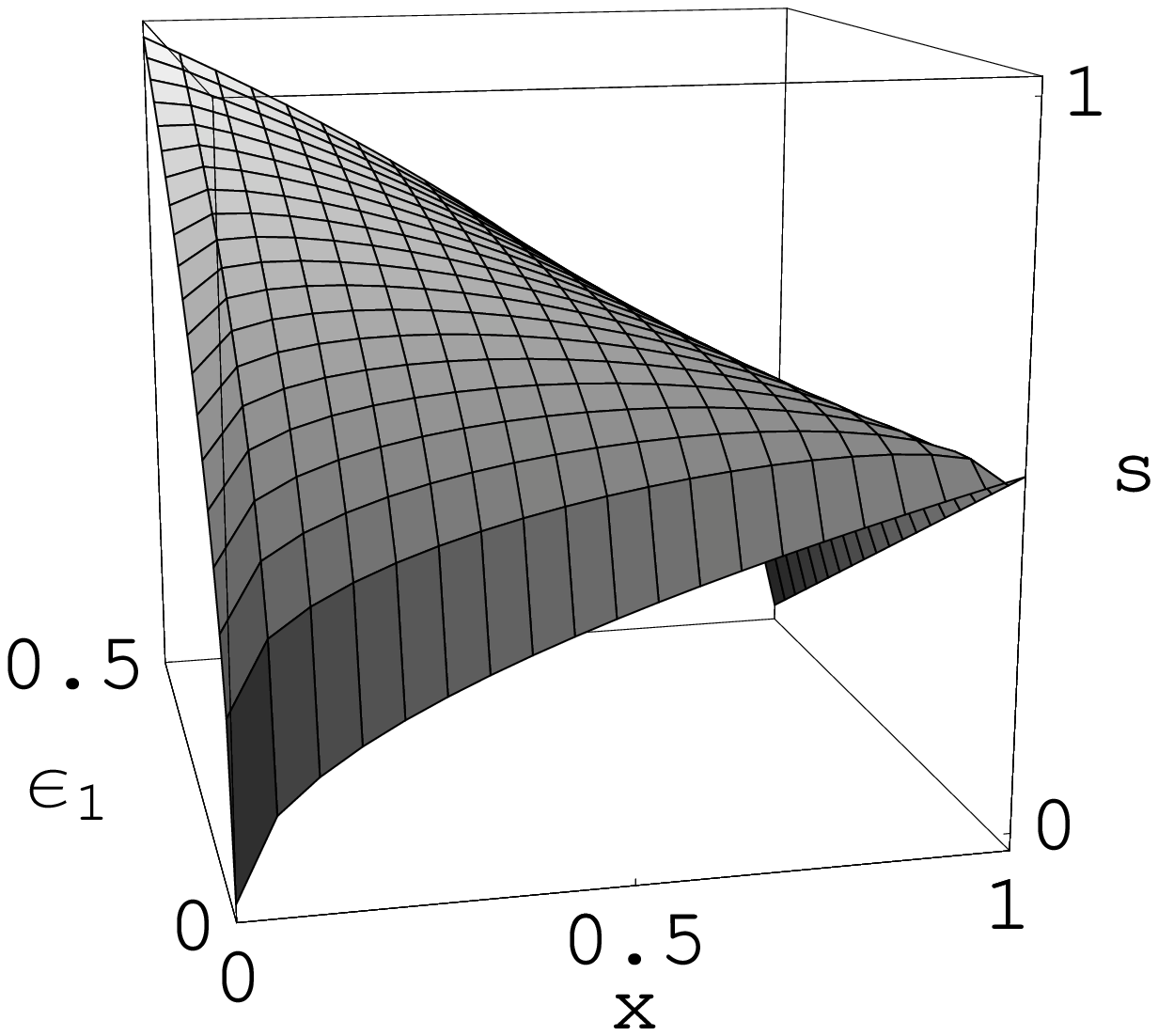}
   \hspace{1truecm}
    \includegraphics[width=2in]{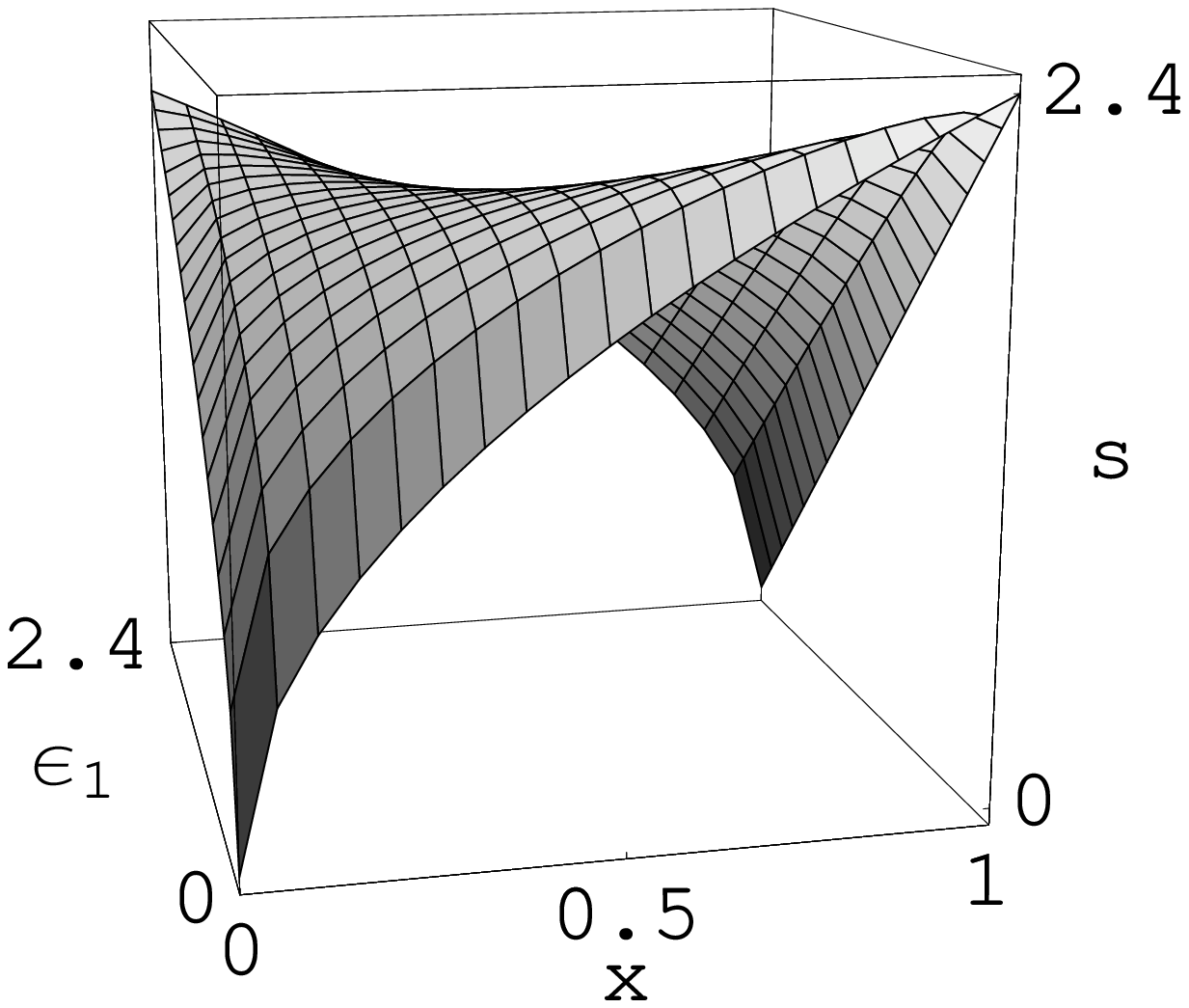}
   \end{center}
   \vspace{.2truecm}
   \hspace{4.6truecm}(a)\hspace{5.7truecm}(b)
   \caption{(a) The entropy is plotted on the vertical axis, as a function of $x, \epsilon_1$, for $\epsilon=0.5$, a value below the point where the three phase structure appears; there is only one maximum at the boundary of the parameter space \quad (b) The entropy graph for $\epsilon=2.4$, a value at which there are three phases; there are two maxima at the boundary of the parameter space and a saddle point in the middle. }
   \label{fig:Entropy_Plot}
\end{figure}

Thus adopting possibility (b) we write the entropy as
\bea
S&=&2\sqrt{2}\pi\sqrt{N-\delta N}\sqrt{E_1\over  m_p}+2\pi\sqrt{\delta N}{E-E_1\over\sqrt{m_km_p}}
\nonumber\\
&=&{2\pi\sqrt{m_k}\over \sqrt{m_p}}\Bigl[\sqrt{2}\sqrt{N-\delta N} 
\sqrt{E\over m_k} + \sqrt{N} {E-E_1\over  m_k}\Bigr]
\label{major}
\eea
If we introduce the dimensionless quantities  
\be
\epsilon={E\over m_k}\,,\quad 
\epsilon_1={E_1\over  m_k}\,,\quad x={\delta N\over N}
\label{microscales}
\ee
we can rewrite the entropy as
\be
S={2\pi\sqrt{N}\sqrt{m_k}\over \sqrt{m_p}}\hat{s}[\epsilon_1,x;\epsilon]
\label{rescaleds}
\ee
where
\be
\hat{s}[\epsilon_1,x;\epsilon]=\sqrt{2}\sqrt{1-x}\sqrt{\epsilon_1}+\sqrt{x}(\epsilon-\epsilon_1)
\label{sxx}
\ee
where for a given $\epsilon$ we should extremize in $\epsilon_1, x$. 

Let us now study the properties of the function $\hat{s}$ for different $\epsilon$.
The extrema of $\epsilon$ occur at points where
\be
{\partial \hat{s}\over \partial \epsilon_1}=0\,,\quad {\partial \hat{s}\over \partial x}=0
\ee
There is only one such point:
\be
\epsilon_1 = \epsilon-1\,,\quad x={1\over 2\epsilon-1}
\ee
This point lies inside the allowed region $\epsilon_1\in[0,\epsilon]$, $x\in[0,1]$ for
\be
\epsilon \ge 1
\ee
The matrix of second derivatives at the extremal point is
\be
{\partial^2 \hat{s}\over \partial (x, \epsilon_1)}=-{1\over \epsilon-1}
\pmatrix{{(2\epsilon-1)^{5/2}\over 8}&{(2\epsilon-1)^{3/2}\over 4}\cr
{(2\epsilon-1)^{3/2}\over 4}&{1\over 2(2\epsilon-1)^{1/2}}}
\ee
Its determinant is
\be
\mathrm{det} {\partial^2 \hat{s}\over \partial (x,\epsilon_1)}=-{(2\epsilon-1)^2\over 8 (\epsilon-1)}
\ee
which is negative for $\epsilon>1$. Thus the extremal point is a saddle point, which is a maximum
along some direction and a minimum along another direction. The value of $\hat{s}$ at the extremum point is
\be
\hat{s}_c=\sqrt{2\epsilon-1}
\ee 

Since the extremum we have found is a saddle, the actual maxima of the entropy must occur at the boundary of the allowed parameter space. We plot $\hat s$ against $x, \epsilon_1$ in fig.(\ref{fig:Entropy_Plot}). When $\epsilon<1$, as in fig.(\ref{fig:Entropy_Plot}) (a), there is one maximum at the boundary
\be
\epsilon_1=\epsilon\,,\quad x=0\,\,\Rightarrow\,\, \hat{s}_a=\sqrt{2\epsilon}
\label{eqfirst}
\ee
This corresponds to having all the energy and entropy in the mode of type (\ref{en1p}) (we call this mode (a)). When $\epsilon>1$ we have the situation in fig.(\ref{fig:Entropy_Plot}) (b), where there are two maxima at the two ends. One as in (\ref{eqfirst}), and the other at
\be
\epsilon_1=0\,,\quad x=1\,\,\Rightarrow\,\, \hat{s}_b=\epsilon
\ee
This latter extremum corresponds to having  all the energy and entropy in a mode of type (\ref{en2p}) (we call this mode (b)). 

Both maxima $\hat{s}_a$ and $\hat{s}_b$ are always greater than $\hat{s}_c$.

To summarize,  we have the following situation:

\bigskip

(1) For $\epsilon<1$, there is only one maximum corresponding to the `black hole phase', phase (a).

\medskip

(2) For $\epsilon>1$ we have three phases, that we interpret as folllows:

\bigskip

(a) black hole phase: $\epsilon_1=\epsilon$, $x=0$, $\hat{s}_a=
\sqrt{2\epsilon}$;

\medskip

(b) uniform black string phase: $\epsilon_1=0$, $x=1$, $\hat{s}_b=\epsilon$;

\medskip

(c) non-uniform black string phase: 
$\epsilon_1=\epsilon-1$, $x=1/(2\epsilon-1)$, $\hat{s}_c=\sqrt{2\epsilon-1}$.

\bigskip

Phase (c) is always unstable.
\subsection{Computing the tension}

The energy above extremality comes from  $P,\bar P$  and  $KK,\overline{KK}$  pairs
\be
E=m_p(n_p+\bar n_p)+m_k(n_k+\bar n_k)=2m_p n_p+2 m_k n_k
\ee
Note that  $m_k$ depends on $L$ but $m_p$ does not. Thus the tension comes only from the excitations of type (b), where both P and KK pairs are excited. Since
$m_k$ depends quadratically on $L$ we have
\be
{\p m_k\over \p L}=2 {m_k\over L}
\label{forpp}
\ee
Thus we get
\be
{\cal T}={\p E\over \p L}={\p E_b\over \p L}=(n_k+\bar n_k) {\p m_k\over \p L}=4 n_k {m_k\over L}
\ee
Using (\ref{formula}) we get\footnote{It is easy to check that the  same result is obtained if we start with the definition
$\mathcal{T}=\Big ({\partial E\over \partial L}\Big )_{S, Q_i}$
and compute the requited variation of $E$ keeping $S, Q_i$ fixed.}
\be
{\cal T}=4 {E-E_b\over 4 m_k} {m_k\over L}={E-E_b\over L}
\ee
The rescaled tension, defined in (\ref{defr}), is
\be
r =  {\mathcal{T} L\over E}={E-E_1\over E}={\epsilon-\epsilon_1\over \epsilon}
\ee

In the uniform black string phase $\epsilon_1=0$ and thus the rescaled tension is
\be
r_b= 1
\ee
In the black hole phase $\epsilon-\epsilon_1=0$ and the tension vanishes
\be
r_a = 0
\ee
In the non-uniform string phase $\epsilon-\epsilon_1=1$ and our microscopic prediction for the
rescaled tension is
\be
r_c={1\over \epsilon}
\label{rc}
\ee

Note that the value $r_a=1$ we have obtained for the uniform string tension is the one expected from
gravity (a fact that had already been noted in \cite{hms}). This can be seen as follows. The rescaled
tension of the neutral system in the uniform string phase is obtained by setting $c_z=0$ in (\ref{neutralmass}), and
it is given by
\be
n_\mathrm{us}={1\over 2}
\ee
Substituting this value in (\ref{n2r}) gives the rescaled tension for the charged uniform string:
\be
r_\mathrm{us}=1
\ee
We thus see that $r_a=r_\mathrm{us}=1$.

\subsection{Comparison between gravity and the microscopic model}
In this subsection we compare various results from gravity with the predictions following from 
our simple microscopic model. It will be convenient to introduce rescaled energies as
\be
\mu\equiv {16 \pi G^{(5)}_N\over L^2} M\,,\quad \epsilon\equiv {16 \pi G^{(5)}_N\over L^2} E
\ee
Using the fact that
\be
{L^2\over 16\pi G_N^{(5)}}={R_z^2 R_y V\over g^2\alpha'^4}=m_k
\ee
we see that the quantity $\epsilon$ defined here coincides with the one introduced in (\ref{microscales}).

\subsubsection{Mapping from the charged system to the neutral one}

\begin{figure}[t] 
   \centering
   \includegraphics[width=3in]{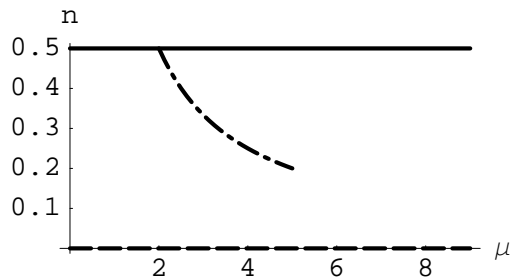} 
   \caption{The phase diagram predicted by the leading order microscopic model. The solid line is the uniform black string. The black hole branch is a horizonal (dashed) line that overlaps with the x-axis -- the tension is zero. The dot-dashed line is the non-uniform string branch.}
   \label{fig:4DnVSmu_micro}
\end{figure}

\begin{figure}[t] 
   \centering
   \includegraphics[width=3in]{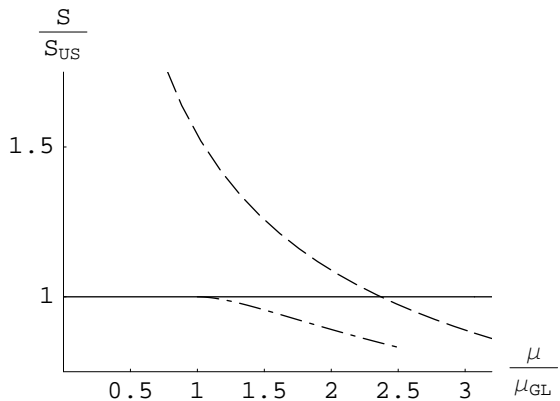} 
   \caption{The entropy graphs for the three phases, with the same coding for the lines as in fig.(\ref{fig:4DnVSmu_micro}).}
  \label{fig:4DsVSmu_micro}
\end{figure}

We would now like to compare the graphs for tension and entropy derived from model (b) with 
the corresponding graphs obtained from gravity  (shown in Figs. \ref{fig:nVSmu_gravity}, \ref{fig:sVSmu_gravity}). The graphs obtained from the gravity computation describe the neutral system, while the results from microscopics are for the charged system. We will thus map the variables in the microscopic computation to the corresponding neutral system, and then plot the results. 

The relations required to effect the needed map were given in section \ref{2chargesection}. Eq.(\ref{n2r}) relates the relative tension $r$ of the charged case to the relative tension $n$ of the neutral case. Eq.(\ref{m2e}) reads as $\epsilon=\mu{1+n\over 3}$ in scaled variables, and relates the energy above extremality of the charged case to the mass in  the neutral case. The entropy  of the charged case is $S'$, but it is more convenient to look at $\hat S$ defined in (\ref{rescaledsgrav}), which is related to the quantities of the neutral case by (\ref{s2shat}). It turns out that $\hat S$ exactly equals the quantity $\hat s$ that we have used in our microscopic analysis. Checking this needs the exact constants relating the $Q_i$ to the corresponding $n_i$, but even without doing this we note that the ratio between $\hat S$ and $\hat s$ is a constant, and this constant will cancel when we plot ratios of entropies.

Carrying out this map to the neutral system, we find the following for the three phases:
\bigskip

(a) black hole phase: 
\be
n=0\,,\quad {S\over S_\mathrm{us}}={4\over \mu^2} \Bigl({2\over 3}\mu\Bigr)^{3/2}
\ee

(b) uniform string phase:
\be
n={1\over 2}\,,\quad {S\over S_\mathrm{us}}=1
\ee

(c) non-uniform string phase:
\be
n={1\over \mu}\,,\quad {S\over S_\mathrm{us}}={4\over \mu^2}\Bigl({2\mu-1\over 3}\Bigr)^{3/2}
\ee

\subsubsection{Tension and entropy graphs}

With the above map to the neutral system we plot the tension and entropy graphs obtained from the microscopic model in figs.(\ref{fig:4DnVSmu_micro}),(\ref{fig:4DsVSmu_micro}) respectively. We should compare these graphs to the graphs of fig.(\ref{fig:nVSmu_gravity}) (b) and fig.(\ref{fig:sVSmu_gravity}) (b) respectively.
The first feature we observe is that the microscopic model exhibits three phases, and the tension and entropies of these phases are in the correct order when compared to the gravity graphs; i.e., the tension of the non-uniform string branch is lower than the tension of the uniform string and higher than the tension of the black hole, and the entropy of the non-uniform string is always lower than the entropy of the other two phases.

We will now compare the microscopic results to the gravity results in more detail. We note that the microscopic model gives vanishing tension for the black hole phase while the gravity computations do not; we will discuss this in the next section. 

\subsubsection{The Gregory-Laflamme point}

Let the Gregory Laflamme (GL) point in the phase diagram be the point where the non-uniform string branch starts off from the uniform string branch. In the microscopic model this point is at
\be
\epsilon_{GL}=1, ~~~r_{GL}=1
\ee
Mapping this to the neutral system we get
\be
\mu_{GL}=2 \epsilon_{GL} = 2
\label{mugl2}
\ee

\subsubsection{Transition energy}
On the gravity side, the energy at which the GL transition happens has been determined numerically in \cite{gubser,wiseman,sorkin}. The value of this energy depends on the number
of non-compact dimensions, and for $d=4$ non-compact dimensions it is given by
\be
\mu_{GL}\equiv {16\pi G_N^{(5)}\over L^2 }M_{GL}\approx 3.52
\ee
Since the microscopic model gives $\mu_{GL}=2$, we find
\be
{\mu_{GL}^{(\mathrm{grav})}\over \mu_{GL}^{(\mathrm{micro})}}=1.76
\ee

Thus we do not get perfect agreement on the GL point in our leading order microscopic model.
But we will see below that once we scale energies so that they are all compared to the GL point, we get good agreement for the tension graph of the non-uniform string. 

\subsubsection{Non-uniform string tension near GL}
In \cite{harmarkobersIII}, from an analysis of the $d=5$ numerical solution of \cite{wiseman}, 
it was found that the following relation describes with good accuracy the non-uniform string phase near to 
the GL point:
\be
{T S\over T_{GL} S_{GL}}-1=x(d)\Bigl({M\over M_{GL}}-1\Bigr)
\ee 
Let us assume that an analogous relation is valid for a generic dimension $d$, with a dimension-dependent
costant $x(d)$. One can combine this relation with the Smarr identity\footnote{In the second equation in (\ref{smarrGL})
we have used the value of the uniform string tension in $d$-dimensions: $n_\mathrm{us}=1/(d-2)$.}
\be
TS={d-2-n\over d-1}M\,,\quad T_{GL} S_{GL} = {d-3\over d-2} M_{GL}
\label{smarrGL}
\ee
to obtain a relation for the rescaled tension 
\be
n = {(d-1)(d-3)\over d-2}(x(d)-1){M_{GL}\over M}+ {(d-2)^2-x(d)(d-1)(d-3)\over d-2}
\label{nneargl}
\ee
The $M$-independent term on the r.h.s. of the above equation vanishes if
\be
x(d)=\bar x(d)={(d-2)^2\over (d-1)(d-3)}
\label{xd}
\ee

For $d=5$, the numerically derived value for the constant $x$ is
\be
x(5)=1.12
\ee
We note that this number is very close to
\be
\bar x(5)={9\over 8}=1.125
\ee 
Thus for $d=5$ the M-independent term in (\ref{nneargl}) vanishes with good accuracy.
We do not know the numerical value for $x(4)$. It seems however reasonable to conjecture that the M-independent
term vanishes also for $d=4$. This happens if
\be
x(4)=\bar x(4)={4\over 3}\approx 1.33
\ee 
We can support this conjecture by expanding (\ref{nneargl}) in powers of $M-M_{GL}$: the first term in this expansion
is known, for $d=4$, from the work of \cite{gubser}:
\be
n={1\over 2}-\hat\gamma (\mu-\mu_{GL})+O((\mu-\mu_{GL})^2)\,,\quad \hat\gamma=.14
\label{nexp}
\ee
Expanding (\ref{nneargl}) we find, for $d=4$
\be
n={3\over 2}(x(4)-1){M_{GL}\over M}+2-{3\over 2}x(4)={1\over 2}-{3\over 2}(x(4)-1){\mu-\mu_{GL}\over \mu_{GL}}+
O((\mu-\mu_{GL})^2)
\label{n1}
\ee
Comparison of (\ref{nexp}) and (\ref{n1}) gives
\be
\hat\gamma={3\over 2}{x(4)-1\over \mu_{GL}}\qquad \Rightarrow\qquad x(4)={2\over 3}\hat\gamma\mu_{GL}+1\approx 1.33
\ee
This confirms, with good accuracy, that $x(4)$ is such that the M-independent term in (\ref{nneargl})
vanishes. If we use $x(4)=\bar x(4)$ in (\ref{nneargl}), we then have
\be
n = {1\over 2} {M_{GL}\over M}={1\over 2} {\mu_{GL}\over \mu}
\label{n2}
\ee 
In the microscopic model we have found that for the non-uniform string
$n={1\over \mu}$. Noting that  $\mu_{GL}=2$ in this model (eq.(\ref{mugl2})), we can write for the microscopic model
\be
n={1\over 2}{\mu_{GL}\over \mu}
\ee
which is in perfect agreement with (\ref{n2}).

\subsubsection{The Smarr relation}

On the gravity side all black objects satisfy the Smarr relation. In the microscopic model,  the  black hole and uniform black string phases correspond to the 4+1 dimensional  black hole and the 3+1 dimensional black hole respectively, and the entropies $S(E, Q_i)$  in each case are known to agree with the gravity computation.  Thus the Smarr relation is satisfied for these two phases. 

For the non-uniform string, the microscopic model gives
\be
S_\mathrm{n-us}(E,N)=S_\mathrm{us}(E_1,N_1)+S_\mathrm{bh}(E_2,N_2)
\ee
where $E=E_1+E_2$, $N=n_1 n_5 = N_1 + N_2$. 

It is easy to check that if a system is made of two non-interacting subsystems, i.e.
\be
E=E_1+E_2, ~~Q_i=Q_{i,1}+Q_{i,2}, ~~S=S_1+S_2
\label{mix}
\ee
and if each subsystem satisfies the Smarr relation, then the total system will satisfy the Smarr relation. Thus it is consistent to identify the saddle point of the microscopic model with the non-uniform string black string.

Even though satisfying the Smarr relation appears trivial in this leading order microscopic model, it will be nontrivial if we look at higher order corrections where we need to  add interactions that link the two different modes of excitation. We will then not have (\ref{mix}), and the interactions will have to be chosen such that the Smarr relation continues to be satisfied. Note that a generic field theory model will not satisfy the required Smarr relation. The dimension $d$ of the spacetime appears in (\ref{smarrnewp}), and this dimension is in general different from the dimension of the dual field theory. Thus the Smarr relation appears to be nontrivial from the viewpoint of the microscopic model. 

\subsection{Other models of fractionation}

At the start of section \ref{microsectionmodel} we had noted three different possible models of fractionation. We analyzed model (b) above in detail; here we note what happens if we look at models (a) and (c).

\subsubsection{Model (a)}

\begin{figure}[t] 
   \centering
   \includegraphics[width=3in]{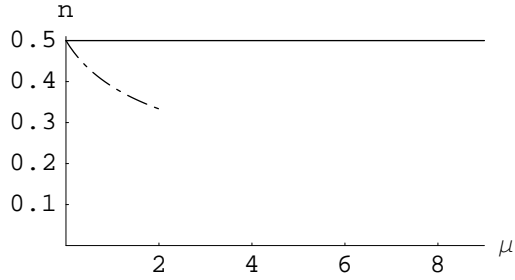} 
   \caption{The tension graph for model (a). Note that the non-uniform string branch starts off at $\mu=0$, which is not in accord with the gravity diagrams (fig.\ref{fig:nVSmu_gravity}). }
    \label{fig:4DsVSmu_micro_alternate_a}
\end{figure}

The analog of (\ref{sxx}) is 
\be
\hat s=\sqrt{2}(1-x)\sqrt{\epsilon_1}+x(\epsilon-\epsilon_1)
\ee
Extremising in $x, \epsilon_1$ gives the point
\be
x={1\over \sqrt{1+2\epsilon}}, ~~~~\epsilon_1={1\over 2}(\sqrt{1+2\epsilon}-1)^2
\label{sol22}
\ee
The determinant of the matrix of second derivatives is
\be
\mathrm{det}{\partial^2 \hat{s}\over \partial (x,\epsilon_1)}=-\Bigl(1+{1\over \sqrt{2 \epsilon_1}}\Bigr)^2
\ee 
Since this is negative, we have a saddle point, and the actual maxima are again at the endpoints of the domain of parameters.
The black hole branch is at $(x=0, \epsilon_1=\epsilon)$, and the uniform black string branch is at $(x=1, \epsilon_1=0)$. But note that the
`non-uniform black string branch', given by (\ref{sol22}), is present for {\it all} values of $\epsilon$. The tension versus mass graph is given in Fig.\ref{fig:4DsVSmu_micro_alternate_a}. We see that the point where the non-uniform black string branch meets the uniform black string branch is at $\epsilon=0$, rather than at some positive value of $\epsilon$. Thus the phase diagram has a qualitative difference from the gravity diagram.

\subsubsection{Model (c)}

\begin{figure}[t] 
   \centering
   \includegraphics[width=3in]{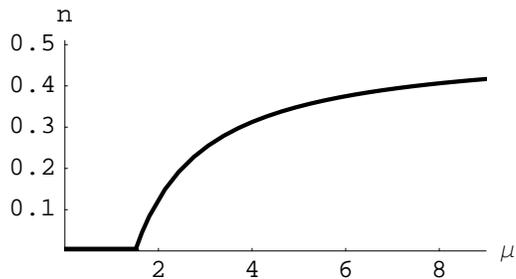} 
   \caption{The phase diagram for model (c). There is only one phase for each $\mu$, which does not agree with
   the gravity diagrams (fig.\ref{fig:nVSmu_gravity}).}
    \label{fig:4DsVSmu_micro_alternate_c}
\end{figure}

In this case the  analog of (\ref{sxx}) is 
\be
\hat  s=\sqrt{2}\sqrt{\epsilon_1}+(\epsilon-\epsilon_1)
\ee
Setting ${\p \hat s\over \p \epsilon_1}=0$ gives
\be
\epsilon_1={1\over 2}
\ee
and ${\p^2\hat s\over \p \epsilon_1^2}=-1< 0$. Thus this point is a maximum when it can be attained.
Thus for $\epsilon<{1\over 2}$ the maximum is at $\epsilon_1=\epsilon$, while for $\epsilon>{1\over 2}$ the maximum is at $\epsilon_1={1\over 2}$. The phase diagram is sketched in Fig.\ref{fig:4DsVSmu_micro_alternate_c}. We see that there is only one phase for all $\epsilon$,  so we have a qualitative difference from the gravity picture.

\subsection{A physical picture of model (b)}

Having seen above that models (a),(c) give phase diagrams that are less in accord with gravity than model (b), let us look at why a model like (b) is suggested by what we know about microscopics. Consider first the case where we have the compactification $M_{4,1}\times T^4\times S^1$, and let there be just one large charge. We let this charge be $n_5$ units of D5; the excitations can then be (fractional) pairs of D1 and P charges.

First suppose that only one kind of charge pairs is excited. We can let this be $P\bar P$. But this situation is very similar to exciting $P\bar P$ pairs on a bound state of D1 branes; the D5 wrapped on $T^4$ acts like an effective string in 6-D and the $P\bar P$ excitations run on this effective string. The picture of this bound state is simple: the strands of the effective string join together to make one `long string', and the $P\bar P$ excitations are transverse vibrations of this long string. The vibrations cause the strands of the long string to separate from each other and spread over some transverse region.

Now consider the excitation mode where two kinds of charge pairs are excited. This situation was studied in \cite{malda}, and it was found that the entropy was exactly reproduced by assuming that the energy of excitation is carried by a `fractional tension D1 brane' living in the D5 worldvolume; the tension of the D1 is fractionated by the factor $1/n_5$. The string excitations represent both $P\bar P$ pairs and $D1\overline {D1}$ pairs. The entropy of such a string is given by  (\ref{en24}). It seems plausible that to get the required fractionation of the tension the D5 branes must be all `stuck together', and the D1 should be  `dissolved' in this D5 bound state to give rise to the effective
string with  fractional tension.

Thus we see that in the mode where one kind of charge pairs are excited  the D5 branes {\it separate} from each other in carrying the excitation, while in the mode where  we have two kinds of excitations we need the D5 branes to be {\it together}. If we are in the process of transition from one mode to the other, then we will have some excitation in each of these modes. But this suggests that we will need to partition the D5 branes into two sets: a fraction $(1-x)$ which will contribute to the first mode and a fraction $x$ which will contribute to the second mode.

Now return to our actual problem, where the compactification is $M_{3,1}\times T^4\times S^1\times \t S^1$ and  instead of the D5 charge we have D1-D5 charge. When we have the mode with only one kind of charge pair the entropy is given by (\ref{en1p}), where the effective string has winding $n_1n_5$ instead of $n_5$.  When we have two kinds of charge pairs, we just note that by dualities we can map this to near extremal D5-KK, which was also studied in \cite{malda}. It was found that we again get a fractional tension effective string living in the worldvolume made by the  D5 and KK branes; all one has to do is replace $n_5$ with $n_5n_k$, which in our duality frame corresponds to replacing $n_5$ everywhere with $n_1n_5$. 

Thus extending our intuition from the D5 charge case to the D1-D5 charge case   we expect that in  the phase transition we will need to partition the product $n_1n_5$ into two parts: a fraction $(1-x)$ for the first mode and a fraction $x$ for the second mode. This is just model (b). 

\subsection{Microscopic model at large momentum}

So far we have added only two large charges, D1, D5, to bring the system near to extremality. But we can also add a large P charge by another boost. In this subsection we consider how the black hole -- black string phase transition will look in the near extremal 3-charge system.

The transition happens when the horizon radius parameter $r_0$ becomes comparable to the compactification length $L$.
We have now chosen all the charges D1,D5,P large, so the charge radii for these charges are all much larger than $L$. We can thus assume that we are looking at a situation with three nonzero charges in 3+1 noncompact dimensions, with a small amount of nonextremality. Since the 4 possible charges for this compactification can be freely permuted by dualities, we can equally well  look at a system where the nonzero charges are D1,D5,KK, and the nonextremality then creates pairs of fractional $P\bar P$ excitations. 

This system has been studied before. If the numbers of the three nonzero charges are $n_1, n_2, n_3$, then we get an effective string with total winding number $n_1n_2n_3$. The kind of entropy that we will encounter here was studied in  \cite{emission}.\footnote{A review is given in \cite{review2}.} If the system is extremal, then we get entropy from partitioning the effective string in all possible ways into component strings, getting an entropy $S=2\pi\sqrt{n_1n_2n_3}$.
If there is some nonextremal energy then pairs of $P\bar P$ excitations can be created on this effective string in the following way. A fraction $f$ of the effective string winding creates entropy as before by breaking up into component strings of different length; this gives an entropy $S=2\pi\sqrt{n_1n_2n_3f}$. The remainder of the effective string joins up into one long component string (thus having no entropy of its own) but this enables the $P\bar P$ excitations  to be fractionated in units of $1/(n_1n_2n_3(1-f))$. This gives an entropy $4\pi\sqrt{n_1n_2n_3(1-f)}\sqrt{E\over 2 m_4}$, by (\ref{formulap}). We then extremize over $f$ to find out how the effective string will partition itself into its two kinds of roles. 

Let us apply this 3-charge model to our present problem, calling the result  `model A'. 
The entropy is thus
\be
S=2\pi\sqrt{f \tilde N}+ 4\pi\sqrt{(1-f)\tilde N}\sqrt{E\over 2 m_k}=2\pi\sqrt{\tilde N}
(\sqrt{f}+\sqrt{2 (1-f) \epsilon})
\ee
where $\t N=n_1n_2n_3$, and we have again defined $\epsilon=E/m_4=E/m_k$ (we have reverted to our duality frame where the nonzero charges are D1, D5, P and the excitations are pairs of KK). The entropy $S$ has only one
extremum, as a function of $f$, given by 
\be
{1\over \sqrt{f}}-{\sqrt{2 \epsilon}\over \sqrt{1-f}}=0\,\,\quad\Rightarrow\,\,\quad f={1\over 2\epsilon+1}
\ee
It is easy to check that this extremum is a maximum. Thus we find a unique 
maximum for every value of the energy: According to model A
the complicated phase structure we have seen emerging in section \ref{microsectionmodel} disappears, at large P,
and we are left with a unique stable phase for every $E$. 
The value of the entropy at the maximum is
\be
S_{max}=2\pi\sqrt{\tilde N}\sqrt{2\epsilon+1}=2\pi\sqrt{\tilde N}
\sqrt{2{E\over m_k}+1}
\label{smax}
\ee
We note that this entropy agrees with the extremal 3-charge black hole entropy \cite{sv}
\be
S_\mathrm{bh}=2\pi\sqrt{\tilde N}
\ee
for small energies ($\epsilon\ll 1$), and with the near-extremal 3-charge
uniform black-string entropy (eq.(\ref{formulap}))
\be
S_\mathrm{us}=2\pi\sqrt{\tilde N}\sqrt{2\epsilon}
\ee
for large energies above extremality ($\epsilon\gg 1$).
We can also compute the value of the rescaled tension $r$ predicted by this 
model. The rescaled tension is defined as
\be
r={L\mathcal{T}\over E}={L\over E}\Big ({\partial E\over \partial L}\Big )_{S, Q_i}
\ee
where $E$ is the energy above 
extremality (so that the total energy is given by 
$M'=Q_{1}+Q_{5}+Q_p+E$).  
Consider the entropy given by (\ref{smax}). Since $m_k\sim L^2$, we find that $S$ is kept constant if the variations of $E,L$ satisfy 
\be
\delta E= 2 E {\delta L\over L}
\ee
Thus
\be
r= {L\over E}{2 E\over L}=2
\label{forpq}
\ee
This agrees with the result from the gravity computation when we have the extremal three large charges limit. In \cite{hkor} it was shown  that in this limit the $r$ vs. $\epsilon$ graphs for the three phases in the gravity computation all degenerate\footnote{This can be seen as follows. For the 3-charge system the energy $M'$  and the charges are given by
\bea
M'&=&M\Bigl(1+{2-n\over 3}(\sinh^2\alpha_1+\sinh^2\alpha_5+\sinh^2\alpha_p)\Bigr)\nonumber\\
Q_i &=& M {2-n\over 3} \sinh\alpha_i\cosh\alpha_i\,,\quad i=1,5,p
\eea
Taking all the charges to be large ($\alpha_i\gg 1$) one finds
\be
M'=Q_1+Q_5+Q_p+M {n\over 2}\,\,\Rightarrow\,\, E = M {n\over 2}
\ee
Since the absolute tension of the neutral system is the same of the one of the
3-charge system, 
the rescaled tension is
\be
r = {M\over E} n = 2
\ee} to the line $r=2$. Thus, as far as the 
tension is concerned, model A gives a prediction which is consistent with gravity.
 
But this analysis may also look a little puzzling, for the following reason. In the gravity computation the three phases collapse to one horizontal line in the $r-\epsilon$ diagram, but there should still be three overlapping phases for each $\epsilon$ in the range where the non-uniform string exists. On the other hand the analysis of model A gives a unique maximum for the entropy for all $\epsilon$, so there seems to be only one phase. 

To emphasize the puzzle still further consider the model (\ref{sxx}) that we have analyzed in the case of two large charges D1, D5, and where we did get three phases for a range of $\epsilon$. In this model we can let $n_p\ne \bar n_p$, thus getting a nonzero P charge. All the results derived above for this model  can be extended to the nonzero $P$ case by changing $n_p, \bar n_p$ in the following way to account for the P charge:
\be
n_p\to n_p e^{\sigma}\,,\quad \bar n_p \to \bar n_p  e^{-\sigma}
\label{pmap}
\ee
The limit of large P is given by taking $\sigma\gg 1$. Let us denote the microscopic model obtained in this way as model B. 
From the point of view of model B, it is clear
that there is a one to one map between the phase diagram we have found at $P=0$ and the one at $P\gg 1$,
given by (\ref{pmap}). In particular model B still predicts three phases (which all degenerate to the line $r=2$ as $\sigma\r\infty$). 

We thus need to ask: If we go slightly off extremality, how does the single line in the tension graph of model A split into three closely spaced lines? We expect that we will see this split if we take into account interactions in the CFT which describes the near extremal 3-charge system. This is a general feature of adding charges to a system; we simplify the system by going closer to extremality, but we may need to take into account interactions between the degrees of freedom of the system  to recover all the physics. We will discuss this issue in more detail in the next section when we talk about the need for interactions to understand the tension in the black hole phase.
 
\section{Tension in the black hole phase}
\setcounter{equation}{0}

Our microscopic model gives a phase diagram that has several features of the phase diagram obtained by numerically solving the equations of general relativity. But one important  feature that is missing is the tension of the black hole branch: the microscopic model gives ${\cal T}=0$ while the numerical results show a tension rising from zero as the mass is increased. This tension was recently related  to  microscopics in \cite{hkor}. In this section we discuss why this tension did not show up in our calculation of the last section, and extract some properties of the tension from microscopic arguments.

\subsection{The source of tension in the black hole phase}

\subsubsection{The leading order gravity calulation}

We first recall the origin of this tension in the gravity picture. A similar calculation can be found for instance in \cite{gorkol2}.

\begin{figure}[htbp] 
\vspace{.6truecm}
\begin{center}
   \includegraphics[width=3in]{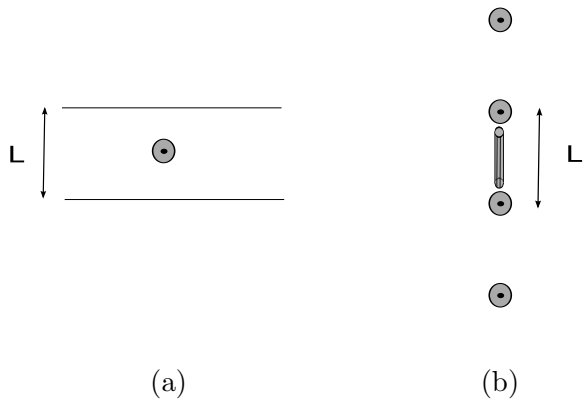} 
   \end{center}
   \vspace{.2truecm}
   \hspace{6truecm}(a)\hspace{3.9truecm}(b)
   \caption{(a) A small black hole in a space with a compact transverse circle \quad (b) Lifting to the covering space replaces the compactification by an infinite  set of images of the hole. Gravitational attraction between the images is given by graviton exchange, which is a one loop of open strings in the dual channel, with all modes of the open string contributing.}
   \label{fig:SqueezedBH}
\end{figure}

Fig.\ref{fig:SqueezedBH}(a) shows a black hole  with radius much smaller than the compactification length. In Fig.\ref{fig:SqueezedBH}(b) we lift to the covering space, getting an infinite  set of images of the black hole. Each copy of the hole feels the attraction of all other copies, and this creates a negative potential energy which is the effect of the compactification. At leading order we can just use the Newtonian potential to compute this energy. In 4+1 dimensions the gravitational potential energy between two bodies of mass $M_1,M_2$ at distance $r$ is
\be
V=-{4 G^{(5)}_N\over 3\pi}{M_1M_2\over r^2}
\ee
The energy between a hole and all its images is
\be
V_{array}=-{4\over 3\pi}{G_N^{(5)}M^2\over L^2}\,2 \sum_{n=1}^{\infty}{1\over n^2}=-{4\pi\over 9}{G_N^{(5)}M^2\over L^2}
\ee
where we have used that
\be
\sum_{n=1}^\infty{1\over n^2}={\pi^2\over 6}
\ee
The part of this energy that we must attribute to the compactification length $L$ is
\be
V_{int}={V_{array}\over 2}= -{2\pi\over 9}{G_{N}^{(5)} M^2\over L^2}
\label{vint}
\ee
The tension is given by ${\cal T}=({\p E\over \p L})_S$. Since the black hole horizon is not distorted to leading order, the entropy is independent of $L$, and we can just compute the $L$ derivative of $V_{int}$ to obtain
\be
L{\cal T}\approx{4\pi\over 9}{G_{N}^{(5)} M^2\over L^2}
\label{tension}
\ee
where the approximation symbol denotes the fact that we have just computed the interaction energy at leading order in the size of the black hole.

The above calculation was for neutral holes, but we can obtain $V_{int}$ for charged holes by relating them to neutral holes via the relations of section \ref{boostsec}. 
We need large D1,D5 charges, and finite P charge. 
 Let $E_0$ be the energy above extremality in the
absence of the interaction $V_{int}$. With no interaction the tension $n$ vanishes, so that 
at leading order one has
\be
E_0 \approx  {M\over 3}\cosh 2\alpha_p
\label{e0}
\ee
The tension is invariant under the addition of charges (eq. (\ref{tensiongr})), so we just get (\ref{tension}) with $M$ obtained through (\ref{e0})
\be
L{\cal T}\approx {4\pi}{G_{N}^{(5)} \over L^2}\Bigl({E_0\over \cosh\alpha_p}\Bigr)^2
\label{tensionone}
\ee
The momentum charge carried by the hole is
\be
Q_p\equiv m_p P \approx {M\over 3}\sinh 2\alpha_p
\label{p0}
\ee
(here $P$ is the integer charge).

 We can write for the charged hole 
\be
{\cal T}=\Bigl({\p E\over \p L}\Bigr)_{Q,S}={\p V_{int}\over \p L}
\ee
Note that the entropy changes in a known way under addition of charge by boosting ($S\r S \cosh\alpha$). From this relation and (\ref{p0}) we see that (to the order we are working at) we keep   $Q_p,S$ fixed if we keep  $M,\alpha_p$ fixed. Thus the tension is related to $V_{int}$ only through variations of $L$, and we find for the charged case a relation analogous to (\ref{vint})
\be
V_{int}\approx -2\pi {G_N^{(5)}\over L^2}\Bigl({E_0\over \cosh 2\alpha_p}\Bigr)^2
\label{vintcharged}
\ee

\subsubsection{The effect in the dual field theory}

Let us now ask how the energy (\ref{vintcharged}) would be obtained in a field theory calculation. We have seen that this energy is obtained from the long range gravitational attraction between the hole and its images. The exchanged graviton can be seen in a dual channel as a 1-loop open string diagram, but note that since $L\gg \sqrt{\alpha'}$ all modes of the open string contribute in the loop integral, not just the lowest energy modes.

\begin{figure}[htbp] 
   \begin{center}
   \includegraphics[width=4in]{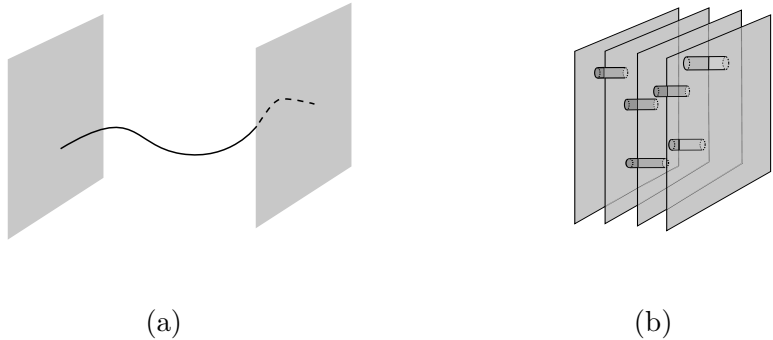} 
   \end{center}
   \vspace{.3truecm}
   \hspace{4.7truecm}(a)\hspace{6truecm}(b)
   \caption{(a) A string stretched between two well separated branes has a closely spaced tower of excitations \quad (b) When a large brane charge is added to the system we only get the ground states of open strings, but the interactions between these encode the original tower of excitations.}
   \label{fig:IntrBranes}
\end{figure}

Now let us ask how this interaction would be seen in a dual field theory description. First look at a simpler system: an open string stretched between two D3-branes,  with the length of this string being $L\gg\sqrt{\alpha'}$; this is pictured in Fig.\ref{fig:IntrBranes}(a). The low energy dynamics of this open string has a tower of closely spaced vibration modes. Now consider such a string, still stretched between two D3 branes, 
in a space with large D3 brane charge -- i.e., in $AdS_5$.  In the gravity picture, the string still has a tower of oscillation modes. But in the dual CFT the open string is a gauge boson, Higgsed because of the separation between the D3 branes at its ends. At weak coupling this gauge boson has no tower of excitation modes. But at larger coupling, we can sum a set of ladder diagrams and see that the tower of oscillation modes is reproduced \cite{klebanovmalda}. Thus we see that the tower of modes in the gravity picture of fig.\ref{fig:IntrBranes}(a) is reproduced in the CFT by degrees of freedom that arise from only the {\it ground} states of open strings; the price we pay is that we have to consider all orders of interaction between these open strings. 

Let us draw lessons from this example for our own system. In the gravity picture fig.\ref{fig:SqueezedBH}(b) we have all modes of the open string stretched between images of the black hole.  The dual CFT in our case is complicated, but can be assumed to arise in some way analogous to the field theory of open strings on branes. To see the vibration modes of the open string in the CFT we thus expect that we will need to consider all orders of interaction between the basic degrees of freedom of the CFT. 

But in our microscopic computation of the phase transition we considered only the free CFT. This should explain why we did not see the tension in the black hole phase.

Interactions in the CFT are difficult to study. In the D1-D5 CFT they involve the insertion of a `twist' operator in  the orbifold CFT \cite{lmorb}. With the presence of the KK-monopole degrees of freedom, we will get other interactions as well. If we do not study these interactions, how can we proceed to study the phase transition?

In \cite{costaperry} a method was developed to get an indirect picture of the CFT effects, and this method was used further in \cite{hkor}. We will recall this approach below, and study it a further to extract some properties of the effective interactions in the CFT.

\subsection{Microscopics}

First consider the black hole in the limit $L\r\infty$. Then we have just a  black hole carrying D1,D5 charges in 4+1 dimensions, and this is described by the  1+1 dimensional CFT. 
 The energy above extremality and the P charge are given by 
\be
E_0 = m_p (n_p+{\bar n}_p)\,,\quad P = n_p-{\bar n}_p
\label{ep}
\ee
Using (\ref{e0}) and (\ref{p0}) we find
\be
n_p={M e^{2\alpha_p}\over 6 m_p}\,,\quad {\bar n}_p = {M e^{-2\alpha_p}\over 6 m_p}
\label{npnpbar}
\ee

In \cite{costaperry,hkor} a `mixed' approach was taken. Suppose that for  $L\r\infty$ the black hole non-extremal energy is $E_0$ and the charge is $P$;  we can find $n_p, \bar n_p$ using (\ref{ep}).  For finite $L$ there is a lowering of the energy due to the interaction with images,  by  an amount $V_{int}$. Thus for the same total energy, there will be higher values of $n_p, \bar n_p$, and a correspondingly higher entropy. This increase in the entropy can be checked against the gravity prediction. This is a `mixed' approach in the sense that $V_{int}$ is computed from gravity, and then fed into the microscopic computation through the values of $n_p, \bar n_p$. 

We will also follow such a   mixed approach, but proceed in a slightly different way that will allow us to better extract some quantities of interest. At $L\r\infty$ the near horizon region of the  D1-D5 system is dual  to a 1+1 dimensional CFT, in which the left and right movers do not interact; thus there is no term coupling $n_p$ to $ \bar n_p$. Because the AdS region ends and goes over to flat space, there is a breaking of conformal symmetry and  
a coupling of left and right movers to the graviton field; this cubic interaction was responsible for Hawking emission from the near-extremal D1-D5 system \cite{dm1}. In the present problem we will let the charge radii of the D1,D5 charges  be much larger than $L$, so this changeover from AdS to flat space is not relevant, and the above coupling is not the leading order interaction between left and right movers. But what we do have is a compact $z$ direction, and KK monopole pairs can wrap around this circle. When  we integrate out the effect
of these pairs we will be left with an interaction between left and right movers. We will now try to understand these effective interactions
in more detail, obtaining information about them  from gravity, the Smarr scaling relation, the behavior  under boost etc.
 
We see from (\ref{vintcharged}), using (\ref{npnpbar}), that the interaction energy 
can be written as
\be
V_{int}= -{8 \pi} {G_N^{(5)} m_p^2\over L^2} n_p {\bar n}_p
\ee
We observe that the interaction energy vanishes in the BPS case ($n_p=0$ or
$\bar n_p=0$), as expected.
The total energy is (to this order)
\be
E = E_0+V_{int}= m_p (n_p+\bar n_p)-{8 \pi} {G_N^{(5)} m_p^2\over L^2} n_p {\bar n}_p
\label{e20}
\ee
This relation gives the energy of the microscopic model to second order in $n_p, \bar n_p$. 

The horizon is not deformed to leading order, so we still have
\be
S=2\pi\sqrt{n_1 n_5}(\sqrt{n_p}+\sqrt{\bar n_p})
\label{entropy}
\ee

How do we continue the study of interactions to higher order? In the $L\r\infty$ limit we had a free CFT and we determined 
$n_p, \bar n_p$ from the relations (\ref{ep}). In the interacting theory there will be an ambiguity in
defining $n_p, \bar n_p$ for a given state. From the gravity perspective we can split the corrections to the
thermodynamics into two kinds of terms: those that come from nonlinearlities of general relativity
and those that arise because the horizon of the black hole is distorted by the compactification. The latter set of terms
do not start till order $r_0^{10}$, where $r_0$ is the horizon radius. The former set of terms includes the interaction 
energy $V_{int}$ derived above as the leading correction, and other similar terms that follow at higher order,
all in integer powers of $r_0^2$ \cite{rothstein}.

We will define $n_p, \bar n_p$ at all orders in the interaction through the relations 
\be
P={Q_p\over m_p}=n_p-\bar n_p\,,\qquad S=2\pi\sqrt{n_1 n_5}(\sqrt{n_p}+\sqrt{\bar n_p})
\label{relations}
\ee
In view of the comments above, this definition is natural to all orders where the horizon is not distorted
(since the internal degrees of freedom giving the entropy $S$ are not affected), and at higher orders there
is no particularly natural scheme to define $n_p, \bar n_p$, so we might as well use (\ref{relations}). The nontrivial information
about the system will then be in the expression for $E$ in terms of $n_p, \bar n_p$.

Note that a priori we have $E=E(S,Q,L)$. Roughly speaking, the Smarr scaling relation will set the $L$ dependence, and
the behavior under boosts will set the $Q$ dependence. Thus the freedom in $E$ will be equivalent to `one function of one variable'. The coefficients describing this remaining function we will take from the gravity calculation -- in principle they can all be determined numerically, but of course we will be looking at only the first few coefficients, and these can be understood analytically.

\subsection{Interactions at second order}

This section is somewhat long, so we first summarize the steps in our computation. In section (\ref{gravhole}) we recall the relations in gravity between the neutral system and the charged one, but in a notation specially adapted to the study of the black hole branch. In section (\ref{defdim}) we introduce dimensionless variables to simplify notation. In section (\ref{smarrhole}) we argue that because of the Smarr relation, we can write the dimensionless energy $\epsilon$ as a function only of rescaled versions of the parameters $n_p, \bar n_p$ (these rescaled parameters are called $\nu, \bar \nu$). In section (\ref{secondhole}) we derive our main result: by using the fact that the tension ${\cal T}$ must remain unchanged under boosting, we derive the term in $\epsilon(\nu, \bar \nu)$ which takes us to one order higher in $\nu, \bar\nu$ than the expression (\ref{e20}). In section (\ref{largenu}) we extend this result to the case where we keep the first order in $\bar\nu$, but all orders in $\nu$. 

\subsubsection{Gravity variables for the small black hole limit}
\label{gravhole}
Let us recall the relation (\ref{erelation})
\be
E={TS\over 2}\cosh 2\alpha_p +{{\cal T}L\over 2}
\label{erelationq}
\ee
Since we are working in the phase where we have a small black hole, it will be convenient to use variables that relate directly to the structure of the hole. The neutral hole in 4+1 dimensions has the metric
\be
ds^2=-(1-{r_0^2\over r^2})dt^2+{dr^2\over 1-{r_0^2\over r^2}}+r^2d\Omega_3^2
\label{metricsymm}
\ee
We have
\be
M={3\pi\over 8 G_N^{(5)}}r_0^2, ~~~S={\pi^2\over 2 G_N^{(5)}}r_0^3, ~~~T={1\over 2\pi r_0}, ~~~TS={\pi\over 4 G_N^{(5)}} r_0^2
\label{mts}
\ee
After we compactify the transverse circle $z$ the metric will no longer have the spherically symmetric form (\ref{metricsymm}).
But we will use the last of the relations (\ref{mts}) to motivate the following definition of $r_0$ even after compactification
\be
TS\equiv {\pi\over 4 G_N^{(5)}} r_0^2
\ee
Let us substitute this in (\ref{erelationq}). We find
\be
E={\pi \over 4 G_N^{(5)}} {r_0^2\over 2} \cosh 2\alpha_p +{{\cal T} L\over 2}
\label{etwo}
\ee
The tension ${\cal T}$ to leading order has been found in (\ref{tensionone}). Putting this in (\ref{etwo}) we get
\be
E\approx {\pi \over 4 G_N^{(5)}} {r_0^2\over 2} \cosh 2\alpha_p+{2\pi G_N^{(5)}\over L^2} \Bigl({M\over 3}\Bigr)^2
\label{ethree}
\ee
The Smarr relation for the neutral system gives
\be
M {2-n\over 3} =TS\equiv {\pi\over 4 G_N^{(5)}} r_0^2
\label{formulaqq}
\ee
When $L\r \infty$ we have $n\r 0$, so for small black holes we have
\be
{M\over 3}\approx {\pi\over 8 G_N^{(5)}} r_0^2
\ee
Substituting this is (\ref{ethree}) we get
\be
E={\pi \over 4 G_N^{(5)}} {r_0^2\over 2} \cosh 2\alpha_p+{2\pi G_N^{(5)}\over L^2} \Bigl({\pi\over 8 G_N^{(5)}}\Bigr)^2 r_0^4+O(r_0^6)
\label{efour}
\ee

\subsubsection{Defining dimensionless variables}
\label{defdim}
The above expression can be simplified by factoring out constants to define dimensionless variables. Thus write
\be
\epsilon={16\pi G_N^{(5)}\over L^2} E, ~~~q={16\pi G_N^{(5)}\over L^2} Q_p, ~~~\rho_0={2\pi\over L} r_0
\ee
Then (\ref{efour}) becomes
\be
\epsilon={\rho_0^2\over 2}\cosh 2\alpha_p+{\rho_0^4\over 32}+O(\rho_0^6)
\label{eeten}
\ee
and we have (using (\ref{qpcharge}),(\ref{formulaqq}))
\be
q={\rho_0^2\over 2}\sinh 2\alpha_p
\label{eeel}
\ee
We also define scaled microscopic values:
\be
\nu={16\pi G_N^{(5)}\over L^2} m_p n_p, ~~~~~ \bar \nu={16\pi G_N^{(5)}\over L^2} m_p \bar n_p
\label{relationsn}
\ee

\subsubsection{The Smarr relation in microscopic variables}
\label{smarrhole}

We are interested in looking at the general expression for $E$ as a power series in $n_p, \bar n_p$; this expression will encode the effect of interactions in the CFT arising from the compactification of the direction $z$. We will now argue that in terms of the dimensionless variables introduced above, we will have
\be
\epsilon=\epsilon(\nu, \bar \nu)
\label{eenine}
\ee
The argument proceeds in three steps:

\bigskip

(A) \quad First recall eq. (\ref{enew}). For the neutral system, we have 
\be
n=n(M,L)
\label{eetwo}
\ee
Using (\ref{qpcharge}) to express $\alpha_p$, we find that we can write
\be
E=E(M,L,Q_p)
\label{eeone}
\ee
Now consider
\be
S=S_{neutral}\cosh\alpha_1\cosh\alpha_5\cosh\alpha_p
\ee
where
\be
S_{neutral}=S_{neutral}(M,L)
\ee
Let us define
\be
\hat S\equiv {S\over \sqrt{Q_1Q_5}}
\ee
Then we find, for large $D1$ and $D5$ charges, 
\be
\hat S={S_{neutral}\cosh\alpha_1\cosh\alpha_5\cosh\alpha_p\over M{2-n\over 3}\sqrt{\sinh\alpha_1\cosh\alpha_1\sinh\alpha_5\cosh\alpha_5}}\r {S_{neutral}\cosh\alpha_p\over M{2-n\over 3}}
\ee
and we find that we can again write
\be
\hat S=\hat S(M,L,Q_p)
\label{eethree}
\ee
We can now eliminate $M$ between (\ref{eeone}) and (\ref{eethree}) to write
\be
E=E(\hat S, L, Q_p)
\label{eefour}
\ee

\medskip

(B)\quad Now consider the Smarr scaling for the 4+1 dimensional gravity solution. The solution is invariant under
\be
E\r L^2E, ~~Q_1\r L^2 Q_1, ~~Q_5\r L^2 Q_5, ~~Q_p\r L^2 Q_p, ~~S\r L^3 S
\label{eefourbis}
\ee
Note that under this scaling we will have
\be
\hat S\r L\hat S
\ee
Thus we can write (\ref{eefourbis}) as
\be
{E\over L^2}={E\over L^2}\Bigl({\hat S\over L}, {Q_p\over L^2}\Bigr)
\ee
In the above scalings we have regarded $G_N^{(5)}$ as a constant, but since this constant has units, its useful to see how it appears in the physical quantities of interest. All energies appear in the gravity theory through the combinations 
\be
G_N^{(5)}E,  ~~G_N^{(5)}Q_1, ~~G_N^{(5)}Q_5, ~~G_N^{(5)}Q_p
\ee
etc. The entropy appears in the gravity theory through the area of a horizon, i.e. as 
\be
G_N^{(5)} S\sim A, ~~~~\hat S={G_N^{(5)}S\over \sqrt {(G_N^{(5)}Q_1)(G_N^{(5)}Q_5)}}
\ee
We can thus write (\ref{eefour}) as
\be
{G_N^{(5)}E\over L^2}={G_N^{(5)}E\over L^2}\Bigl({\hat S\over L}, {G_N^{(5)}Q_p\over L^2}\Bigr)
\label{eesix}
\ee

\medskip

(C) \quad Now note that
\be
{16\pi G_N^{(5)}\over L^2}Q_p=\nu-\bar \nu
\label{eesev}
\ee
Also,
\be
\hat S={S\over \sqrt{Q_1Q_5}}={2\pi\sqrt{n_1n_5}(\sqrt{n_p}+\sqrt{\bar n_p})\over \sqrt{n_1m_1n_5m_5}}=
{2\pi(\sqrt{n_p}+\sqrt{\bar n_p})\over \sqrt{m_1m_5}}
\ee
where $m_1, m_5, m_p$ will be used to denote the masses of individual quanta of D1, D5, P charges. Now note that
\be
m_1m_5m_p={\pi\over 4 G_N^{(5)}}
\ee
and we find
\be
{\hat S\over L}= \sqrt{\nu}+\sqrt{\bar \nu}
\label{eeeig}
\ee
Returning to (\ref{eesix}) we see that the LHS is just $\epsilon$ (upto a numerical constant), and the arguments on the
RHS are (from (\ref{eesev}),(\ref{eeeig})) just functions of $\nu, \bar \nu$. Thus we establish (\ref{eenine}).

\subsubsection{The second order correction in the microscopic picture}
\label{secondhole}
We have defined the microscopic parameters $n_p, \bar n_p$ through (\ref{relations}), and their rescaled versions $\nu, 
\bar \nu$ through (\ref{relationsn}). We now wish to work out the expansion of $\epsilon(\nu, \bar \nu)$ in powers of its arguments, using what we know from gravity.

At leading order in $\nu, \bar \nu$ we have
\be
\epsilon=\nu+\bar \nu, ~~~~q=\nu-\bar \nu
\ee
We note some other general properties of $\epsilon$. First, it must be symmetric in $\nu, \bar \nu$. Second, if $\bar \nu=0$ then the system is BPS, and the energy is not corrected by the compactification length $L$ being finite. This means that for $\bar \nu=0$ we must have precisely $\epsilon=\nu$; there are thus no terms in $\epsilon$ of type $\nu^2, \nu^3, \dots$, and similarly there are no terms $\bar \nu^2, \bar \nu^3, \dots$. 

The gravity calculation of the next to leading correction to the energy  (eq. (\ref{e20})) is written in dimensionless form as
\be
\epsilon=\nu+\bar \nu-{1\over 2} \nu\bar \nu
\ee
At next order we would have
\be
\epsilon=\nu+\bar \nu-{1\over 2}\nu\bar \nu+\gamma\, \nu\bar \nu(\nu+\bar \nu)
\label{mueq}
\ee
We will now see that we can determine the constant $\gamma$ just by using the fact that the tension does not change under boosting. Note that boosting changes the charge $Q_p$, and will also change $\nu, \bar \nu$. But since the tension is to be left invariant, we get a constraint on  expansion coefficients like $\gamma$ in (\ref{mueq}). Let us carry out this computation. 

The tension is given through
\be
L{\cal T}=L{\p\over \p L}E=L{\p\over \p L} \Bigl[{L^2\over 16\pi G_N^{(5)}}\epsilon(\nu, \bar \nu)\Bigr]
\ee
The $L$ derivative is done by keeping $S$ and $Q_p$ fixed; from (\ref{relations}) this is
equivalent to keep $n_p$ and $\bar n_p$ fixed, and hence $\nu, \bar \nu$ scale as $L^{-2}$. It follows that for a term in $\epsilon$ of the form $\nu^a\bar \nu^b$ we will have the contribution to the tension
\be
\tau\equiv {16\pi G_N^{(5)}\over L^2}L{\cal T}\r -2(a+b-1)\nu^a\bar \nu^b
\ee
Thus for the terms in (\ref{mueq}) we will get
\be
\tau =\nu\bar \nu-4\gamma\, \nu\bar \nu(\nu+\bar \nu)
\label{taurel}
\ee
Let us now convert to gravity variables, using (\ref{eeten}),(\ref{eeel}). We have
\be
\epsilon=\nu+\bar \nu-{1\over 2}\nu\bar \nu={\rho_0^2\over 2}\cosh2\alpha_p+{\rho_0^4\over 32}
\ee
\be
q=\nu-\bar \nu={\rho_0^2\over 2}\sinh 2\alpha_p
\ee
where we have written the $\nu, \bar \nu$ expansion only to the orders that we will need.\footnote{Since
$\nu,\bar \nu\sim \rho_0^2$, if follows from (\ref{taurel}) that to compute $\tau$ up to order $\rho_0^6$ we only need $\epsilon$ and $q$ up to order $\rho_0^4$.} These relations invert to give
\be
\nu={\rho_0^2\over 4}e^{2\alpha_p}+{\rho_0^4\over 32}, ~~~~ \bar \nu={\rho_0^2\over 4}e^{-2\alpha_p}+{\rho_0^4\over 32}
\label{eenn}
\ee
Substituting in (\ref{taurel}) we find
\be
\tau={\rho_0^4\over 16}+{\rho_0^6\over 64} (1-8\gamma) \cosh 2\alpha_p +O(\rho_0^8)
\ee
Requiring that $\tau$ not depend on $\alpha_p$ then gives
\be
\gamma={1\over 8}
\label{gamma}
\ee
and 
\be
\tau={\rho_0^4\over 16} +O(\rho_0^8)
\label{taurelbis}
\ee

The relation (\ref{etwo}) rewritten in rescaled variables appears as
\be
\epsilon={\rho_0^2\over 2}\cosh2\alpha_p + {\tau\over 2}
\ee
Substituting  (\ref{taurelbis}) we get 
\be
\epsilon= {\rho_0^2\over 2}\cosh2\alpha_p+{\rho_0^4\over 32}+ C \rho_0^6+O(\rho_0^8)
\ee
with
\be
C=0
\ee
Thus we find that the coefficient of $\rho_0^6$ in $\epsilon$ vanishes. The determination of this $O(\rho_0^6)$ term is the main result of the computation above. Determination of this term has given us  the next correction to the energy beyond  (\ref{e20}). This term was determined by an explicit and detailed gravity computation in  \cite{cincinnati, rothstein}. It may appear remarkable that we have found it here by a simple  argument involving the invariance of the tension under boosting, and one might wonder where the physics of this correction has been fed into our analysis. This physical input is actually hidden in the assumption that we can expand $\epsilon$ in integer powers of $\nu, \bar \nu$ to the order that we are working.
In other words, the vanishing of the  $O(\rho_0^6)$ term in the energy does not seem to have a simple interpretation in the gravity computation, but it does have a simple interpretation in the microscopic description. Thus we  find that the numbers $n_p \bar n_p$ defined through (\ref{relations}) give a good definition of `quasi-particle excitations' to the first few orders in the interaction, and using an expansion in integral powers of these numbers automatically reproduces the $O(\rho_0^6)$
term in the energy.

\subsubsection{First order in $\bar \nu$, all  orders in $\nu$}
\label{largenu}

In the above we have worked to second order in the excitations $\nu, \bar \nu$ which correspond to left and right movers respectively. We will now extend our microscopic expression for the energy $\epsilon$ to include terms with arbitrary orders in $\nu$, while keeping only the first order in $\bar \nu$. This means that we allow  a large net P charge, since the left moving excitation $\nu$ is no longer small. Our technique will be to again use the fact that the tension ${\cal T}$ does not depend on the boost parameter $\alpha_p$, but now we will take our limits in a slightly different way, as follows.

We will still let $\rho_0$ be our small parameter, but we are interested in large boosts $\alpha_p$. Thus we will let
\be
\rho_0\r 0, ~~~\alpha_p\r\infty, ~~~\rho_0^2 e^{2\alpha_p}\r ~{\rm finite}
\label{limitnew}
\ee
Note that we will thus have
\be
\rho_0^2e^{-2\alpha_p}\sim O(\rho_0^4)
\ee
etc.

We now have
\be
\epsilon = \nu+\bar \nu -{\nu\bar \nu\over 2}I(\nu)\,,\quad q= \nu-\bar \nu
\label{theequations}
\ee 
and
\be
\tau=\nu\bar \nu (I(\nu)+\nu I'(\nu))+O(\bar \nu^2)
\label{taufirst}
\ee
To understand the limits that we are taking, look at the gravity description of these variables.
First consider the expression for energy in the expansion where $\rho_0$ is small, and $\alpha_p$ is finite. From the way that energy transforms under boosting, we know that
\be
\epsilon={\rho_0^2}\sinh^2\alpha_p+\tilde \epsilon_0(\rho_0)
\ee
where $\tilde\epsilon_0(\rho_0)$ is the energy of the system without the P charge; thus $\tilde\epsilon_0(\rho_0)$
does not depend on $\alpha_p$. We can thus write
\be
\epsilon= {\rho_0^2\over 2}\cosh2\alpha_p + {\rho_0^4\over 32}+O(\rho_0^6)\,,\quad 
q = {\rho_0^2\over 2}\sinh2\alpha_p
\label{gravityfirst}
\ee
where we have used $\cosh 2\alpha_p=2\sinh^2\alpha_p+1$ and the terms $O(\rho_0^6)$ do not depend upon $\alpha_p$.
The latter fact then tells us that even when we take the limit (\ref{limitnew}) the terms dropped in $\epsilon$ are $O(\rho_0^6)$. The expression for $q$ is exact.

We now find
\be
q\approx {\rho_0^2\over 4} e^{2\alpha_p}\equiv q_0\,,\quad {\epsilon-q\over 2}\approx 
{\rho_0^2\over 4} e^{-2\alpha_p}\Bigl(1+{q_0\over 4}\Bigr)
\label{eqnsol}
\ee
We will in general keep terms upto order $\rho_0^4$, but we have written the expression for $q$ only to leading order since we will not need higher orders. We invert (\ref{theequations}) to find $\nu, \bar \nu$ to the required order
\be
\nu=q+{\epsilon-q\over 2}\Bigl(1-{q\over 4}I(q)\Bigr)^{-1}+O((\epsilon-q)^2)\,,
\quad 
\bar \nu={\epsilon-q\over 2}\Bigl(1-{q\over 4}I(q)\Bigr)^{-1}+O((\epsilon-q)^2)
\label{nnbarfirst}
\ee
Using  (\ref{eqnsol}) and then substituting in (\ref{taufirst}) we have
\be
\tau={\rho_0^4\over 16}\Bigl(1+{q_0\over 4}\Bigr)
\Bigl(1-{q_0\over 4}I(q_0)\Bigr)^{-1}(I(q_0)+q_0 I'(q_0))+O(\rho_0^8)
\label{tupto8}
\ee
Invariance under boost requires
\be
\Bigl(1+{q_0\over 4}\Bigr)
\Bigl(1-{q_0\over 4}I(q_0)\Bigr)^{-1}(I(q_0)+q_0 I'(q_0))=const.
\ee
Moreover when $q_0=0$ one should recover the previous result (\ref{taurelbis}), so the constant in the above equation should be unity. 
The differential equation
\be
\Bigl(1+{\nu\over 4}\Bigr)
\Bigl(1-{\nu\over 4}I(\nu)\Bigr)^{-1}(I(\nu)+\nu I'(\nu))=1
\ee
with the boundary condition $I(0)=1$, has a unique solution given by
\be
I(\nu)=\Bigl(1+{\nu\over 4}\Bigr)^{-1}
\label{iofn}
\ee
(\ref{iofn}) is the main result of this subsection. Substituting back in $\epsilon$ in (\ref{theequations}) we find that we have
the energy to first order in $\bar n_p$ but to all order in $n_p$. We have again obtained this result by a simple argument; the essential physics here was that $n_p$ can be changed by boosting (which adds P charge), and properties of black objects change in simple ways under boosting.

Note that the first order expansion of $I(\nu)$ in powers of $\nu$ implies
$\gamma={1\over 8}$, in agreement with our earlier result (\ref{gamma}).

\section{Discussion}

Let us summarize the computations of this paper. We have noted that gravity computations reveal an interesting
phase diagram for the black hole -- black string transition, with three phases for a range of $r_0/L$.  We can add charges to the neutral system by `boosting plus dualities'. This does not add or lose any information as far as the gravity computation is concerned, since the neutral and boosted variables are related in a simple and invertible way. But it does map the problem to a near extremal one, where we can look for microscopic models. 

For $r_0/L$ very small  we get the near extremal 2-charge system  in 4+1 noncompact dimensions, and for $r_0/L$ very large we get the near extremal 2-charge system  in 3+1 noncompact dimensions. Both these systems have been given a microscopic description in earlier work on the subject.  We propose a simple model that interpolates between these two microscopic models; the nontrivial step is guessing what kind of `fractionation' will describe the D1-D5 degrees of freedom in the interpolating regime. We find that the three-phase nature of the phase diagram is reproduced if we assume that the the two large charges $n_1, n_5$ bind to make $N=n_1n_5$ units of an effective charge, and this product $N$ is then partitioned between subsystems that behave like one or the other of the two limiting systems mentioned above. We suggest a physical reason why this kind of `fractionation' is reasonable for this problem. Note that even though the microscopic model looks like a superposition of two subsystems (eq.(\ref{major})), we have not just superposed a black hole and a black string; the fact that the {\it product} $N=n_1n_5$ partitions between the two subssytems forces the entire state to be a nontrivial bound state of all the excitations.

The non-uniform string appears as a saddle point between two maxima of the entropy; this accords with the fact that the entropy of this branch is always lower than the entropy of the other two phases, so it is always unstable. The $n-\mu$ graph describing the tension of this branch appears to have a profile $n\sim 1/\mu$ from the gravity computation; the microscopic computation also gives the functional behavior $n\sim 1/\mu$.

The microscopic computation based on such a simple model does not give any tension for the black hole branch. We have argued that this tension will be seen only if we include interactions like the `twist interaction'  that are present in the CFT when we move away from the orbifold point (where we have a `free CFT') to a point in moduli space that has the couplings appropriate to the actual gravity dual. We did not look at such interaction terms directly, but instead  followed a `mixed' approach similar to that in \cite{costaperry,hkor}. Thus  we used the knowledge of the leading order tension in the gravity computation to see what effective interactions we obtain in the CFT. 
We were then able to reproduce the next order term in the tension by using the invariance of the tension under `boosting'. Since boosting adds charge it changes $n_p, \bar n_p$ and the non-extremal energy $E$, and we get a nontrivial condition when we require invariance of the tension ${\cal T}={\p E\over \p L}$. This next order term in the tension is equivalent to a corresponding correction term in the energy, and this correction is seen to agree with detailed computations of the energy carried out in
\cite{cincinnati,rothstein}.  The essential physics we used was to assume that we have an expansion of the energy in integer powers of $n_p, \bar n_p$ to the first few orders; this suggest that the $n_p, \bar n_p$ (defined through (\ref{relations})) are good definitions of `quasi-particles' for the system to this order in the interaction.

While our microscopic model gives some general qualitative agreements with the gravity phase diagram, there are also some differences. First of course is the issue of tension for the black hole branch mentioned above. Next, consider for  example the point where the non-uniform black string branch starts off from the uniform string branch. In the microscopic model this branch starts at a mass value that is $\approx .57$ times  the value seen in the gravity diagram. It would also be nice to see in the microscopic model how the black hole branch bends around to join  with the non-uniform string branch, and nature of the physics and the joining point.

In principle all features of the gravity phase diagram are contained in the correct microscopic dual, so we look upon this phase diagram as a large source of data on the interactions of branes that make up black holes and other black objects. 
The phase diagram offers us continuous curves that the microscopics must fit; further, we are probing a domain where the black object is changing character from one type to another, so we learn about aspects of the CFT that we do not encounter when looking at just near extremal entropy or low energy Hawking radiation. Thus we consider the model proposed here as just a first pass at understanding the microscopics of branes dual to black objects.

\section*{Acknowledgements}

We thank T. Harmark, N. Obers and B. Kol for explaining many aspects of their work to us, and A. Saxena, K. Skenderis and Y. Srivastava for many helpful comments.
This work was supported in part by DOE grant DE-FG02-91ER-40690.

\appendix
\section{Map between neutral and 2-charge hole}
\label{first}

\renewcommand{\theequation}{A.\arabic{equation}}
\setcounter{equation}{0}

In this appendix, starting from the neutral geometry (\ref{neutral}), we generate a geometry 
carrying D1-D5 charges: the D1 extends along the direction $y$ and the D5 along $y$ and $z_a$, with
$a=1,\ldots,4$. Of the remaining $4+1$ directions, one, denoted by $z$, is compactified on a circle 
of length $L$. We are thus left with $d=4$ non-compact directions. 

The desired 2-charge solution is generated by acting on (\ref{neutral}) with the following 
sequence of boosts and dualities
(we denote by $\mathcal{B}_y$ a boost along $y$ and by $T_{ij\ldots}$ a T-duality in the directions 
$i,j,\ldots$)
\bea
&&\!\!\!\!\!\!\!\!\!\!\!\!
\mathrm{neutral}\stackrel{\mathcal{B}_y}{\longrightarrow} P_y 
\stackrel{T_y}{\longrightarrow} NS1_y\stackrel{\mathcal{B}_y}{\longrightarrow} 
NS1_y, P_y\stackrel{S}{\longrightarrow} D1_y, P_y\stackrel{T_{1234}}{\longrightarrow} D5_{y1234}, P_y\nonumber\\
&&\quad\stackrel{S}{\longrightarrow} NS5_{y1234}, P_y\stackrel{T_{y1}}{\longrightarrow} NS5_{y1234},NS1_y\stackrel{S}{\longrightarrow} 
D5_{y1234}, D1_y
\eea 
We list the geometries obtained at each step. The notation is as follows: $ds^2$ denotes the string
metric, $\Phi$ is the dilaton, $B^{(2)}$ the NS-NS B-field and $H^{(3)}$ its field strength, $C^{(p)}$
is the RR $p$-form gauge potential and $G^{(p+1)}$ the corresponding field strength, 
$\star_B$ denotes the Hodge  dual with respect to $ds^2_B$.

\begin{itemize}
\item $\mathcal{B}_y$:
\bea
ds^2 &\!\!=\!\!& -H_5^{-1} U dt^2 + H_5 \Bigl(dy-(H_5^{-1}-1)\coth\alpha_5 \,dt\Bigr)^2+ds^2_B + dz_a dz_a
\nonumber\\
H_5\!\!&=\!\!&  1+(1-U)\sinh^2\alpha_5
\eea

\item $T_y$:
\bea
ds^2 &\!\!=\!\!& H_5^{-1} (-U dt^2+dy^2) +ds^2_B + dz_a dz_a\nonumber\\
B^{(2)}_{ty}&\!\!=\!\!&-(H_5^{-1}-1)\coth\alpha_5\,,\quad e^{2\Phi}=H_5^{-1}
\eea

\item $\mathcal{B}_y$:
\bea
ds^2 &\!\!=\!\!& H_5^{-1} [-H_1^{-1} U dt^2+ H_1(dy-(H_1^{-1}-1)\coth\alpha_1\,dt)^2]\nonumber\\
&& \quad+ds^2_B + dz_a dz_a\nonumber\\
B^{(2)}_{ty}&\!\!=\!\!&-(H_5^{-1}-1)\coth\alpha_5\,,\quad e^{2\Phi}=H_5^{-1}\nonumber\\
H_1&\!\!=\!\!& 1+(1-U)\sinh^2\alpha_1
\eea

\item $S$:
\bea
ds^2 &\!\!=\!\!& H_5^{-1/2} [-H_1^{-1} U dt^2+ H_1(dy-(H_1^{-1}-1)\coth\alpha_1\,dt)^2]\nonumber\\
&&\quad +H_5^{1/2}(ds^2_B + dz_a dz_a)\nonumber\\
C^{(2)}_{ty}&\!\!=\!\!&-(H_5^{-1}-1)\coth\alpha_5\,,\quad e^{2\Phi}=H_5
\eea

\item $T_{1234}$:
\bea
ds^2 &\!\!=\!\!& H_5^{-1/2} [-H_1^{-1} U dt^2+ H_1(dy-(H_1^{-1}-1)\coth\alpha_1\,dt)^2] \nonumber\\
&&\quad+H_5^{1/2}ds^2_B + H_5^{-1/2}dz_a dz_a\nonumber\\
C^{(6)}_{tyz_1\ldots z_4}&\!\!=\!\!&-(H_5^{-1}-1)\coth\alpha_5\,,\quad 
e^{2\Phi}=H_5^{-1}\nonumber\\
G^{(7)}_{rtyz_1\ldots z_4}&\!\!=\!\!& -\partial_r H_5^{-1}\coth\alpha_5\,,\quad G^{(3)}=-U^{-1/2} \partial_r H_5 \coth\alpha_5 \,\star_B dr
\eea

\item $S$:
\bea
ds^2 &\!\!=\!\!& -H_1^{-1} U dt^2+ H_1(dy-(H_1^{-1}-1)\coth\alpha_1\,dt)^2\nonumber\\
&&\quad +H_5 ds^2_B + dz_a dz_a\nonumber\\
 H^{(3)}&\!\!=\!\!&-U^{-1/2} \partial_r H_5 \coth\alpha_5 \,\star_B dr\,,\quad 
e^{2\Phi}=H_5
\eea

\item $T_{y1}$:
\bea
ds^2 &\!\!=\!\!& H_1^{-1}(- U dt^2+ dy^2) +H_5 ds^2_B + dz_a dz_a\nonumber\\
 H^{(3)}&\!\!=\!\!&-U^{-1/2} \partial_r H_5 \coth\alpha_5 \,\star_B dr-\partial_r H_1^{-1}\coth\alpha_1\, dr\wedge dt\wedge dy\nonumber\\
e^{2\Phi}&\!\!=\!\!&{H_5\over H_1}
\eea

\item $S$:
\bea
ds^2 &\!\!=\!\!& (H_5H_1)^{-1/2}(- U dt^2+ dy^2) + (H_5H_1)^{1/2} ds^2_B + 
\Bigl({H_1\over H_5}\Bigr)^{1/2} dz_a dz_a\nonumber\\
 G^{(3)}&\!\!=\!\!&-U^{-1/2} \partial_r H_5 \coth\alpha_1 \,\star_B dr-\partial_r H_1^{-1}\coth\alpha_1\, dr\wedge dt\wedge dy\nonumber\\
e^{2\Phi}&\!\!=\!\!&{H_1\over H_5}
\eea
\end{itemize}

Converting from string to Einstein frame, we find
\bea
ds^2_E &\!\!=\!\!& H_1^{1/4}H_5^{-1/4}[H_1^{-1}(- U dt^2+ dy^2) + H_5 ds^2_B +  dz_a dz_a]\nonumber\\
 G^{(3)}&\!\!=\!\!&-U^{-1/2} \partial_r H_5 \coth\alpha_5 \,\star_B dr-\partial_r H_1^{-1}\coth\alpha_1\, dr\wedge dt\wedge dy\nonumber\\
e^{2\Phi}&\!\!=\!\!&{H_1\over H_5}\,,\quad H_1 = 1+(1-U)\sinh^2\alpha_1\,,\quad H_5 = 1+(1-U)\sinh^2\alpha_5
\label{chargedEApp}
\eea

Energy and tension are derived from the large $r$ behavior of the Einstein metric. 
Assume that the neutral metric has the following asymptotic expansion
\be
 -g_{tt}=1-{c_t\over r}\,,\quad g_{zz}=1+{c_z\over r}\,,\quad  
g_{yy}=1\,,
\quad g_{z_a z_a}= 1\nonumber\\
\ee
Let $M$ and $\mathcal{T}$ be the mass and tension of the neutral system, and define the relative tension
\be
n ={L\mathcal{T}\over M}
\label{relativetens}
\ee
Einstein's equations relate these quantities to the $c_t$ and $c_z$ defined above:
\be
M= {L\over 16\pi G_N^{(5)}} \Omega_2 (2 c_t -c_z)\,,\quad \mathcal{T}={\Omega_2\over 16\pi G_N^{(5)}}(c_t - 2 c_z)\,,\quad n = {c_t - 2 c_z\over 2 c_t -c_z}
\label{neutralmass}
\ee

Let $g'_{\mu\nu}$ denote  the components of the metric in (\ref{chargedEApp}),  $C^{(2)}$ the RR 2-form 
gauge field and  $C^{(6)}$ the dual 6-form field. Their asymptotic 
expansions are  of the form
\bea
&&-g'_{tt}=1-{c'_t\over r}\,,\quad g'_{zz}=1+{c'_z\over r}\,,\quad  
g'_{yy}=1+{c'_y\over r}\,,
\quad g'_{z_a z_a}= 1+{c'_a\over r}\nonumber\\
&&C^{(2)}_{ty}={c'_1\over r}\,,\quad C^{(6)}_{t y z_1\ldots z_4}={c'_5\over r}
\eea
Let $M'$ and $\mathcal{T}'$ be the mass and tension of the charged system and  $Q_1$ and $Q_5$ 
denote the D1 and D5 charges (with a normalization such that $M' = Q_1 + Q_5$
for the extremal hole). The relations linking mass, tension and charges to the asymptotic
fall-off of the metric are 
\bea
&&M' ={L\over 16\pi G_N^{(5)}}  \Omega_2 (2 c'_t -c'_z-c'_y -4 c'_a)\,,\quad \mathcal{T}'=
{\Omega_2\over 16\pi G_N^{(5)}}(c'_t - 2 c'_z -c'_y - 4 c'_a)\nonumber\\
&&Q_1 = {L\over 16\pi G_N^{(5)}} \Omega_2 c'_1\,,\quad Q_5 ={L\over 16\pi G_N^{(5)}}  \Omega_2 c'_5
\label{chargedmass}
\eea

From (\ref{chargedEApp}) we find the relations between the coefficients $c_i$ and $c'_i$:
\bea
c'_t &=& c_t +{3\over 4}c_t \sinh^2\alpha_1 + {1\over 4}c_t \sinh^2\alpha_5\,,\quad 
c'_z = c_z +{1\over 4}c_t \sinh^2\alpha_1 + {3\over 4}c_t \sinh^2\alpha_5\nonumber\\
c'_y &=& -{3\over 4}c_t \sinh^2\alpha_1 - {1\over 4}c_t \sinh^2\alpha_5\,,\quad 
c'_u = {1\over 4}c_t \sinh^2\alpha_1 - {1\over 4}c_t \sinh^2\alpha_5\nonumber\\
c'_1 &=& c_t \sinh\alpha_1\cosh\alpha_1\,,\quad c'_5 = c_t \sinh\alpha_5\cosh\alpha_5
\label{c's}
\eea
We deduce from (\ref{neutralmass}), (\ref{chargedmass}) and (\ref{c's}) that
\be
M' = M\Bigl(1+{2-n\over 3}(\sinh^2\alpha_1+\sinh^2\alpha_5)\Bigr)\,,\quad \mathcal{T}'=\mathcal{T}\,,
\quad Q_i =M {2-n\over 3}\sinh\alpha_i \cosh\alpha_i\,\,(i=1,5)
\label{2charge}
\ee

Let us now consider temperature and entropy. Temperature is proportional to the surface gravity
$\kappa$, which is given by
\be
\kappa^2 = -{1\over 4} [g^{tt}g^{rr}(\partial_r g_{tt})^2]|_{r=r_0}
\label{surfacegravity}
\ee
with $r=r_0$ the location of the horizon. From (\ref{chargedE}), we find that the components of the 
2-charge metric are related as follows to the components of the
neutral metric (\ref{neutral})
\be
g'_{tt}=(H_1^{1/4}H_5^{-1/4})H_1^{-1}g_{tt}\,,\quad g'_{rr}=(H_1^{1/4}H_5^{-1/4})H_5 g_{rr}
\ee 
The surface gravity of the 2-charge system is thus
\bea
\kappa'^2 &=&  -{1\over 4} [g'^{tt}g'^{rr}(\partial_r g'_{tt})^2]|_{r=r_0}\nonumber\\
&=&\kappa^2 (H_1 H_5)^{-1}|_{r=r_0}={\kappa^2\over \cosh^2\alpha_1\cosh^2\alpha_2}
\eea
where we have used the fact that $g_{tt}=0$ at the horizon and that
\be
H_i|_{r=r_0}=\cosh^2\alpha_i\,\,(i=1,5)
\ee 
Then the temperatures of the neutral and charged holes ($T$ and $T'$) are related as
\be
T' = {T\over \cosh\alpha_1\cosh\alpha_5}
\label{temperature}
\ee

Entropy is proportional to the horizon area. Using again (\ref{chargedE}), we see that the 
horizon areas before and after the boost ($A_h$ and $A'_h$) are related as
\be
A'_h = (H_1 H_5)^{1/2} A_h
\ee
and thus the respective entropies ($S$ and $S'$) scale as
\be
S' = S \cosh\alpha_1 \cosh\alpha_5
\label{entropyApp}
\ee
Note that
\be
T S= T' S'
\ee

\end{document}